\documentclass[twocolumn]{aastex63}
\usepackage{newtxtext,newtxmath}

\usepackage{threeparttable, upgreek}
\usepackage{lineno}
\usepackage{color}
\usepackage{xspace}

\newcommand{\LambdaCDM}{$\Lambda$CDM\xspace}
\newcommand{\COBEF}{{\it COBE/FIRAS}\xspace}

\shorttitle{Radio Synchrotron Background Workshop 2 Summary}
\shortauthors{J. Singal, N. Fornengo, M. Regis, et al.}

\begin{document}
\title{THE SECOND RADIO SYNCHROTRON BACKGROUND WORKSHOP:\\ CONFERENCE SUMMARY AND REPORT}

\author{J. Singal}
\affiliation{Department of Physics, University of Richmond, 138 UR Drive, Richmond, VA 23173, USA\email{jsingal@richmond.edu}}

\author{N. Fornengo}
\affiliation{Dipartimento di Fisica, Universit\`{a} di Torino, via P. Giuria 1, I--10125 Torino, Italy}
\affiliation{Istituto Nazionale di Fisica Nucleare, Sezione di Torino, via P. Giuria 1, I--10125 Torino, Italy}
\author{M. Regis}
\affiliation{Dipartimento di Fisica, Universit\`{a} di Torino, via P. Giuria 1, I--10125 Torino, Italy}
\affiliation{Istituto Nazionale di Fisica Nucleare, Sezione di Torino, via P. Giuria 1, I--10125 Torino, Italy}

\author{G. Bernardi}
\affiliation{INAF-Istituto di Radioastronomia, via Gobetti 101, 40129, Bologna, Italy}
\affiliation{Department of Physics and Electronics, Rhodes University, PO Box 94, Makhanda, 6140, South Africa}
\affiliation{South African Radio Astronomy Observatory (SARAO), Black River Park, 2 Fir Street, Observatory, Cape Town, 7925, South Africa}

\author{D. Bordenave}
\affiliation{Department of Astronomy, University of Virginia, 530 McCormick Road, Charlottesville, VA 22904, USA}
\affiliation{Central Development Laboratory, National Radio Astronomy Observatory, 1180 Boxwood Estate Road, Charlottesville, VA 22903, USA}

\author{E. Branchini}
\affiliation{Department of Mathematics and Physics, University Roma Tre, L. San Leonardo Murialdo, 1, Rome, Italy}
\affiliation{Department of Physics, University of Genoa, Via Dodecaneso 33, 16146 Genoa, Italy}

\author{N. Cappelluti}
\affiliation{Department of Physics, University of Miami, Knight Physics Building 330, Coral Gables, FL 33146, USA}

\author{A. Caputo}
\affiliation{Department of Particle Physics, School of Physics and Astronomy, Tel Aviv University, Chaim Levanon St. 55, Tel Aviv-Yafo, Israel}
\affiliation{Department of Particle Physics and Astrophysics, Weizmann Institute of Science, 234 Herzl Street, POB 26, Rehovot Rehovot, Israel}
\affiliation{Department of Theoretical Physics, CERN, Esplanade des Particules 1, P.O. Box 1211, Geneva 23,
Switzerland}

\author{I. P. Carucci}
\affiliation{Dipartimento di Fisica, Universit\`{a} di Torino, via P. Giuria 1, I--10125 Torino, Italy}
\affiliation{Istituto Nazionale di Fisica Nucleare, Sezione di Torino, via P. Giuria 1, I--10125 Torino, Italy}

\author{J. Chluba}
\affiliation{Jodrell Bank Centre for Astrophysics,  Department of Physics \& Astronomy, The University of Manchester, M13 9PL, UK}

\author{A. Cuoco}
\affiliation{Dipartimento di Fisica, Universit\`{a} di Torino, via P. Giuria 1, I--10125 Torino, Italy}
\affiliation{Istituto Nazionale di Fisica Nucleare, Sezione di Torino, via P. Giuria 1, I--10125 Torino, Italy}

\author{C. DiLullo}
\affiliation{NASA Goddard Space Flight Center, 8800 Greenbelt Road, Greenbelt, MD 20771, USA}

\author{A. Fialkov}
\affiliation{Institute of Astronomy, University of Cambridge, Madingley Road, Cambridge CB3 0HA, UK}

\author{C. Hale}
\affiliation{Institute for Astronomy, University of Edinburgh, Blackford Hill, Edinburgh EH9 3HJ, UK}

\author{S. E. Harper}
\affiliation{Jodrell Bank Centre for Astrophysics,  Department of Physics \& Astronomy, The University of Manchester, M13 9PL, UK}

\author{S. Heston}
\affiliation{Department of Physics, Virginia Tech, Blacksburg, VA 24061-0435, USA}

\author{G. Holder}
\affiliation{Department of Physics, University of Illinois, 1110 West Green Street, Urbana, IL 61801, USA}

\author{A. Kogut}
\affiliation{NASA Goddard Space Flight Center, 8800 Greenbelt Road, Greenbelt, MD 20771, USA}

\author{M. G. H. Krause}
\affiliation{Centre for Astrophysics Research, Department of Physics, Astronomy and Mathematics, University of Hertfordshire, Hatfield, Hertfordshire AL10 9AB, UK}

\author{J. P. Leahy}
\affiliation{Jodrell Bank Centre for Astrophysics,  Department of Physics \& Astronomy, The University of Manchester, M13 9PL, UK}

\author{S. Mittal}
\affiliation{Dept. of Theoretical Physics, Tata Institute of 
Fundamental Research, Homi Bhabha Road, Mumbai 400005, India}

\author{R. A. Monsalve}
\affiliation{Space Sciences Laboratory, University of California, 7 Gauss Way, Berkeley, CA 94720, USA}
\affiliation{School of Earth and Space Exploration, Arizona State University, 781 Terrace Mall, Tempe, AZ 85287, USA}
\affiliation{Facultad de Ingenier\'ia, Universidad Cat\'olica de la Sant\'isima Concepci\'on, Alonso de Ribera 2850, Concepci\'on, Chile}

\author{G. Piccirilli}
\affiliation{Department of Physics, University of Roma Tor Vergata, Via della Ricerca Scientifica, 1 00133 Rome, Italy}

\author{E. Pinetti}
\affiliation{Fermi National Accelerator Laboratory, PO Box 500
Batavia, IL 60510, USA}
\affiliation{Kavli Institute for Cosmological Physics, University of Chicago, Chicago, IL 60637, USA}

\author{S. Recchia}
\affiliation{Dipartimento di Fisica, Universit\`{a} di Torino, via P. Giuria 1, I--10125 Torino, Italy}
\affiliation{Istituto Nazionale di Fisica Nucleare, Sezione di Torino, via P. Giuria 1, I--10125 Torino, Italy}

\author{M. Taoso}
\affiliation{Istituto Nazionale di Fisica Nucleare, Sezione di Torino, via P. Giuria 1, I--10125 Torino, Italy}

\author{E. Todarello}
\affiliation{Dipartimento di Fisica, Universit\`{a} di Torino, via P. Giuria 1, I--10125 Torino, Italy}
\affiliation{Istituto Nazionale di Fisica Nucleare, Sezione di Torino, via P. Giuria 1, I--10125 Torino, Italy}

\begin{abstract}
We summarize the second radio synchrotron background workshop, which took place June 15--17, 2022 in Barolo, Italy.  This meeting was convened because available measurements of the diffuse radio zero level continue to suggest that it is several times higher than can be attributed to known Galactic and extragalactic sources and processes, rendering it the least well-understood electromagnetic background at present and a major outstanding question in astrophysics.  The workshop agreed on the next priorities for investigations of this phenomenon, which include searching for evidence of the Radio Sunyaev-Zel'dovich effect, carrying out cross-correlation analyses of radio emission with other tracers, and supporting the completion of the 310~MHz absolutely calibrated sky map project.  
\end{abstract}


\section{Introduction} \label{intro}
\begin{figure}
\includegraphics[width=3.5in,height=3.0in]{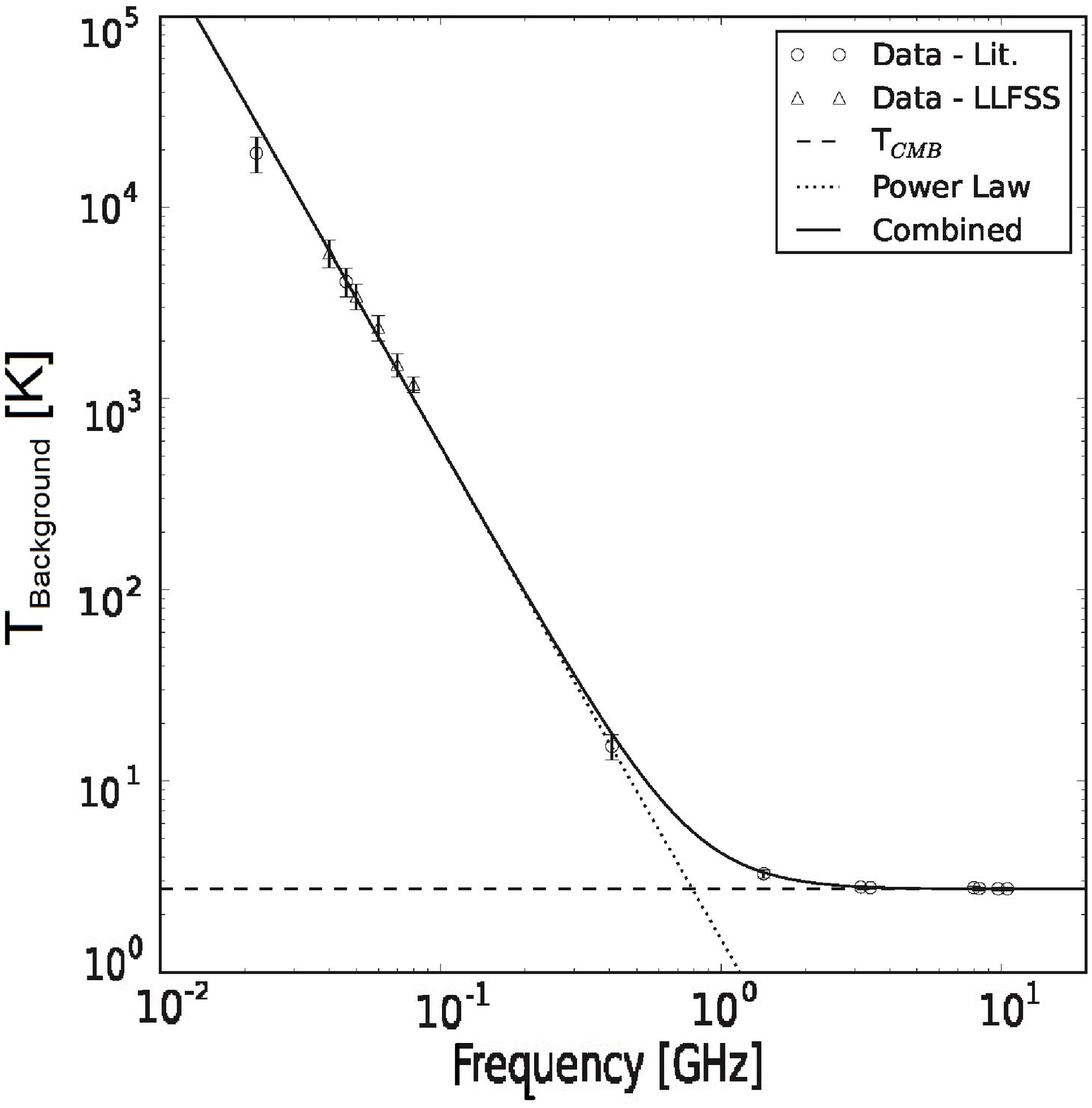}\\
\\
\includegraphics[width=3.5in]{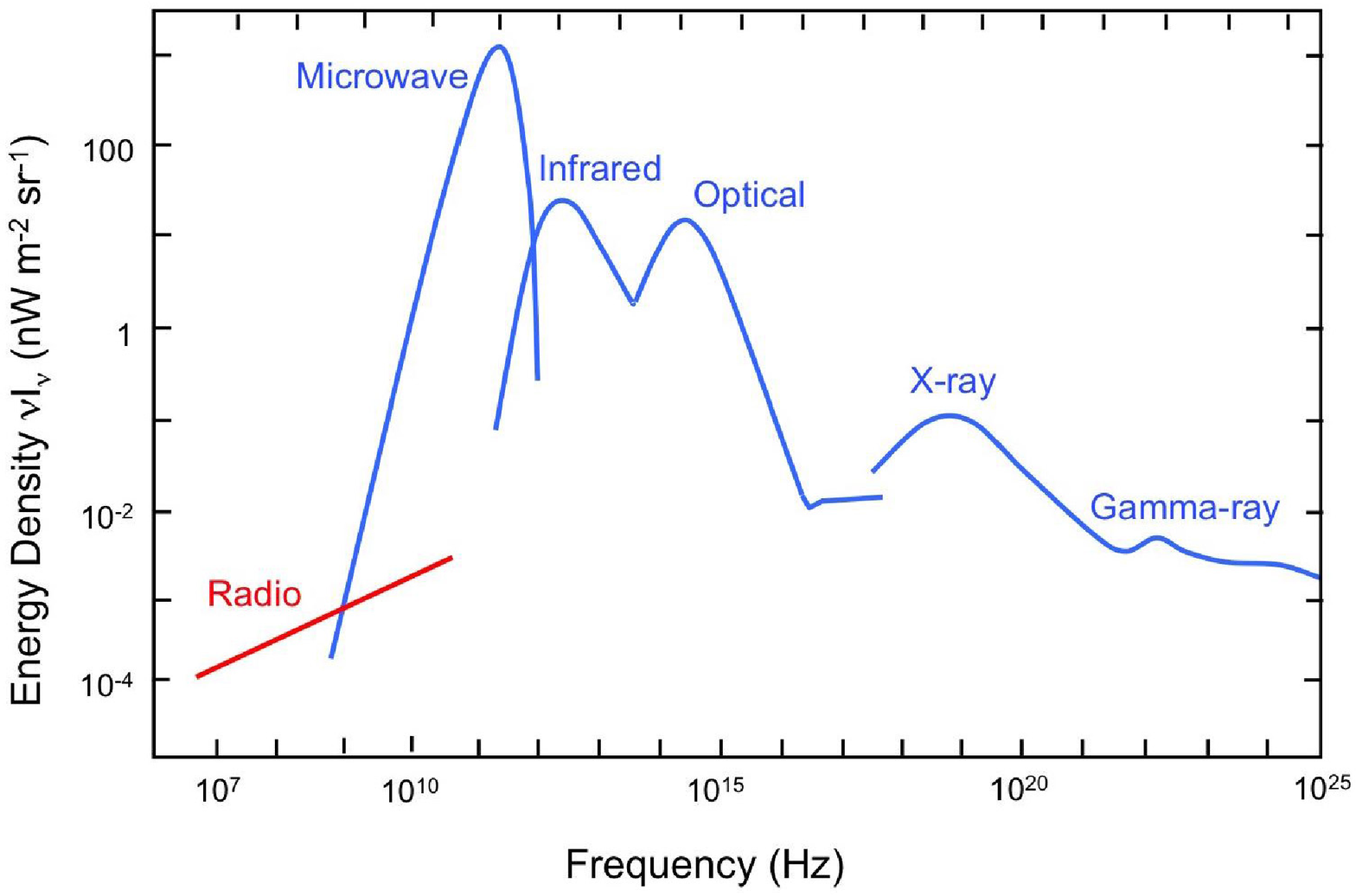}
\caption{{\bf Top}: The radio sky zero level in radiometric temperature units, reproduced from \citet{DT18}, as measured by several different instruments or surveys reporting an absolute zero-level calibration.  The spectrum has power-law at frequencies below $\sim$10 GHz above the CMB level.   Results are shown for ARCADE~2 at 3--10 GHz \citep{Fixsen11}, \citet{RR86} at 1.4 GHz, \citet{Haslam} at 408 MHz,  \citet{Maeda99} at 45 MHz, and \citet{Roger99} at 22 MHz, as well as several points reported by \citet{DT18}.  {\bf Bottom}: The photon backgrounds in the universe in units of spectral energy surface brightness density.  Reproduced from \citet{CP1}.} 
\label{CRB}
\end{figure}

The radio synchrotron background (RSB) is a phenomenon that has been of interest to many in the astrophysical community in recent years.  Combining Absolute Radiometer for Cosmology, Astrophysics, and Diffuse Emission 2 (ARCADE~2) measurements from 3 to 90$\mathrm{~GHz}$ \citep{Fixsen11} with several radio maps at lower frequencies from which an absolute zero level has been inferred  \citep[recently summarized in][]{DT18} reveals a synchrotron background brightness spectrum,
\begin{equation}
T_\mathrm{BGND}(\nu) = 30.4 \pm 2.6 \mathrm{K} \, \left(
\frac{\nu}{310\mathrm{~MHz}}
\right)^{-2.66 \pm 0.04} \, + \, T_{\rm CMB}
\label{T_B}
\end{equation}
where $T_{\rm CMB}$ is the frequency-independent contribution of 2.725~K due to the cosmic microwave background (CMB).  This surface brightness of the radio zero level, shown in Figure \ref{CRB}, is several times higher than that attributable to  known classes of discrete extragalactic radio sources \citep[e.g.][]{Hardcastle21}, which is in dramatic contrast to the observed monopole components at infrared, optical/UV, X-ray, and gamma-ray wavelengths.  

In addition, various Galactic and extragalactic production mechanisms are highly constrained, due to observations of other galaxies and tracers of Galactic radio emission \citep[e.g.][]{RB1}, the radio/far-infrared correlation \citep[e.g.][]{YL12} and inverse-Compton implications for the X-ray background \citep[e.g.][]{Singal10}, effects of the presence of a radio background for \ion{H}{1}~21-cm cosmological signals \citep[e.g.][]{Fialkov_19}, and, potentially, observed arcmintue-scale anisotropy constraints at GHz \citep{Holder14} and MHz \citep{LF21} frequencies.

A second workshop on the RSB was merited, given that it touches on so many contemporary issues in astrophysics, and especially given the developments in both theory and observation that have taken place in the five years since the first radio synchrotron background workshop was held in 2017 at the University of Richmond in Virginia, USA.  That previous workshop has been summarized in \citet{CP1}.  

This report presents a summary of the presentations, discussions, and conclusions of the 2022 workshop for the rest of the astrophysical community.  \S \ref{details} reports the logistical details of the meeting.  \S \ref{overview} gives a brief summary of the problem of the RSB.  \S \ref{summary} gives a summary of individual presentations in the workshop.  The overall conclusions from the various discussions are presented in \S \ref{conc}.

\section{Meeting Details}\label{details}

The organizing committee consisted of Jack Singal (University of Richmond), Marco Regis (University of Torino and INFN) and Nicolao Fornengo (University of Torino and INFN).  This workshop was part of a series of Barolo Astroparticle Meetings (BAM) which are organized by the theoretical astroparticle group of the University of Torino and INFN-Torino on a semiregular basis.  Participation in the workshop was by invitation of the organizing committee only, and it was conducted in-person at the Hotel Barolo in Barolo, Piedmont, Italy.  Individuals who participated in the workshop are listed in Table~\ref{participants}.  Most participants arrived to Barolo on Tuesday, June~14, 2022 and departed on Saturday, June~18, 2022.  The program consisted of presentation and discussion sessions, with the latter featuring both small-group brainstorming and large group time.  The workshop website\footnote{https://agenda.infn.it/event/28184/} contains a repository of the program and many of the presentation slides.  

\begin{table}
\scriptsize
\caption{Participants} 
\label{participants}
\begin{tabular}{ll}
Name & Institution\\
\hline
Gianni Bernardi & INAF-Istituto di Radioastronomia \& Rhodes University \\
David Bordenave & National Radio Astronomy Observatory \& University of Virginia \\
Enzo Branchini & University of Roma Tre \& University of Genoa \\
Nico Cappelluti & University of Miami \\
Andrea Caputo & Tel Aviv University \& Weizmann Institute of Science \& CERN\\
Isabella P. Carucci & University of Torino \& INFN  \\
Jens Chluba & University of Manchester \\
Alessandro Cuocco & University of Torino \& INFN \\
Chris DiLullo & NASA Goddard Space Flight Center \\
Anastasia Fialkov & University of Cambridge \\
Nicolao Fornengo & University of Torino \& INFN \\
Catherine Hale & University of Edinburgh \\
Stuart Harper & University of Manchester \\
Sean Heston & Virginia Tech \\
Gil Holder & University of Illinois at Urbana-Champaign \\
Alan Kogut & NASA Goddard Space Flight Center \\
Martin Krause & University of Hertfordshire \\
Patrick Leahy & University of Manchester \\
Shikhar Mittal & Tata Institute of Fundamental Research \\
Raul Monsalve & UC Berkeley \& Arizona State Univ. \& Universidad Cat\'olica\\
Elena Pinetti & Fermilab and Kavli Institute of Chicago\\
Sarah Recchia & University of Torino \& INFN \\
Marco Regis & University of Torino \& INFN \\
Jack Singal & University of Richmond \\
Marco Taoso & University of Torino \& INFN \\
Elisa Todarello & University of Torino \& INFN \\
\hline
\end{tabular}
\end{table}

\section{Scientific Overview}\label{overview}

An apparent bright high Galactic latitude diffuse radio zero level has been reported since the early era of radio astronomy \citep{WO51,Wyatt53}, into the 1960s \citep{Costain60,Bridle67}, and was seen in data from the 1980s \citep{Phillips81,Beuermann85}.  Early analyses simply assumed that the observed intensity was some mixture of an extragalactic background from radio point sources with the remainder allocated to a Galactic contribution, and neither of these was particularly well constrained at the time.  Renewed interest came with combining the ARCADE~2 balloon-based absolute spectrum data from 3--90~GHz  \citep{Fixsen11,Singal11} with absolutely calibrated zero level single-dish degree-scale-resolution radio surveys at lower frequencies \citep[e.g.][]{Haslam} which agreed on a bright radio synchrotron zero level.  In part as a result of the first RSB workshop, the Long Wavelength Array (LWA1) collaboration calculated an absolute zero-level calibration and measurement of the sky zero-level at 40--80~MHz which agreed with the level established by ARCADE~2 and the single-dish radio surveys \citep{DT18}. Another relevant recent result is the absolute calibration computed for the \citet{guzman2011} 45-MHz and the \citet{landecker1970} 150-MHz all-sky maps, and soon to be computed for the \citet{kriele2022} 159~MHz map, by the EDGES collaboration using measurements from several single-dipole-antenna instruments as discussed in \S \ref{Monsalve} and \citep{monsalve2021}.

The radio synchrotron zero level as reported by ARCADE~2 and lower-frequency surveys is spatially uniform enough to be considered a ``background,'' thus it would join the astrophysical backgrounds known in all other regions of the electromagnetic spectrum.

If it is indeed at the level given by equation (\ref{T_B}), which seems to be overwhelmingly likely, the origin of the radio background would be one of the mysteries of contemporary astrophysics. It is difficult to produce the observed level of surface brightness by known processes without violating existing constraints. A brief review of some recent literature on the subject follows:

Some authors \citep[e.g.][]{SC13} have proposed that the background originates from a radio halo surrounding the Milky Way Galaxy.  There are important difficulties with a Galactic origin, however.  With magnetic field magnitudes determined by radio source rotation measures to be present in the Galactic halo of $\sim$1\,$\mu$G, the same electrons energetic enough to produce the radio synchrotron background at the observed level would also overproduce the observed X-ray background through inverse-Compton emission \citep{Singal10}.  Also the observed correlation between radio emission and that of the singly ionized carbon line (\ion{C}{2}) would imply an overproduction of the observed level of emission from that line above observed levels \citep{Kogut11}.  Furthermore, independent detailed modeling of the structure of the diffuse radio emission at different frequencies does not support such a large halo \citep{Fornengo14}.  A halo of the necessary size and emissivity would make our Galaxy anomalous among nearby similar spiral galaxies \citep{RB1}.  Lastly, analyses of the observed polarization of the synchrotron sky seem to disfavor a Galactic origin, as discussed in \S \ref{Kogut} of this report.

However an extragalactic origin for the RSB also presents many challenges.  Several authors have considered deep radio source counts \citep{Vernstrom11,Vernstrom14,Condon12,Hardcastle21} and concluded that if the RSB surface brightness is produced by discrete extragalactic sources they must be a therefore undetected population that is very low flux and therefore very numerous in number, at least an order of magnitude more numerous than the total number of galaxies in the observable universe.  These results are in agreement with others who have probed whether active galactic nuclei \citep[AGN --][]{Ball2011} or other objects \citep{Singal10} are numerous enough, and are further discussed in \S \ref{Hale} of this paper.  

Works in the literature have noted that if the RSB were produced by sources that follow the known correlation between radio and far-infrared emission in galaxies, the far-infrared background would be overproduced above observed levels  \citep{Ponente11,YL12}, while others have claimed that the correlation may evolve with redshift and have noted the implications for the radio background \citep{IvisonA,IvisonB,Magnelli15}. Other works have investigated the anisotropy power of the RSB which seems to be too low at GHz frequencies \citep{Holder14} to trace the distribution of large-scale structure in the universe while being orders of magnitude higher on the same angular scales at MHz frequencies \citep{LF21}.  Observations have ruled out a large signal from the cosmic filamentary structure \citep{Vernstrom17}.  Other important constraints come from considering that the presence of a significant radio background at the redshifts of reionization could have a dramatic effect on the observed \ion{H}{1} 21-cm absorption trough as discussed in \citet{FH18}, \citet{EW18}, \citet{MF19}, \citet{Fialkov_19}, \citet{Mondal20}, \citet{Nat21}, and \citet{MR22}, and here in \S \ref{Fialkov}.

Such constraints have led various authors to investigate potential origins such as supernovae of massive population~III stars \citep{Biermann14}, emission from Alf\'{e}n reacceleration in merging galaxy clusters \citep{FL15}, an enhancement in the Local Bubble \citep[although see \S \ref{Krause} of this report]{Sun08}, annihilating dark matter (DM) in halos or filaments \citep{Fornengo11,Hooper14,FL14,Fortes19} or ultracompact halos \citep{Yang13}, ``dark'' stars in the early universe \citep{Spolyar09}, dense nuggets of quarks \citep{LZ13}, injections from other potential particle processes as discussed in \citet{CV14} and \citet{Pospelov} and here in \S \ref{Chluba} and \S \ref{Caputo}, and accretion onto primordial black holes (PBHs) as discussed here in \S \ref{Cappeluti} and \S \ref{Mittal}.

\section{Summary of Individual Presentations}\label{summary}

\subsection{Introduction and New Measurements --- Jack Singal}
\label{Singal}

This talk presented a brief summary of the points of \S \ref{overview} and then introduced two new measurements of relevance to the RSB.  These are the 310~MHz absolute map to be made with the Green Bank Telescope and custom instrumentation, and the anisotropy power spectrum measurement at 140~MHz made with LOw Frequency ARray \citep[LOFAR --][]{VH13} observations.  The former is in preparation while the latter has been completed.

\begin{figure}
\includegraphics[width=2.5in]{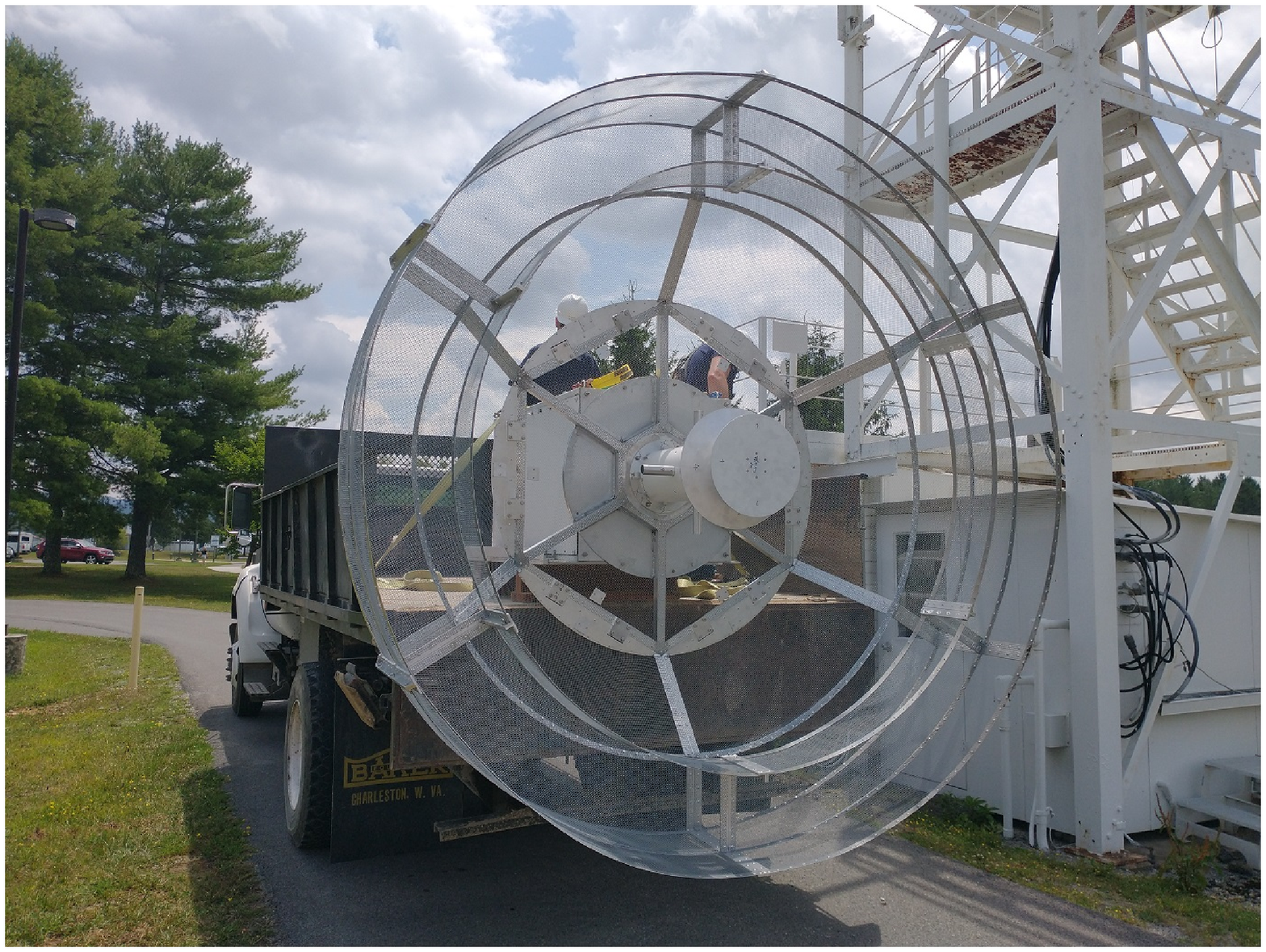}\\
\includegraphics[width=2.5in]{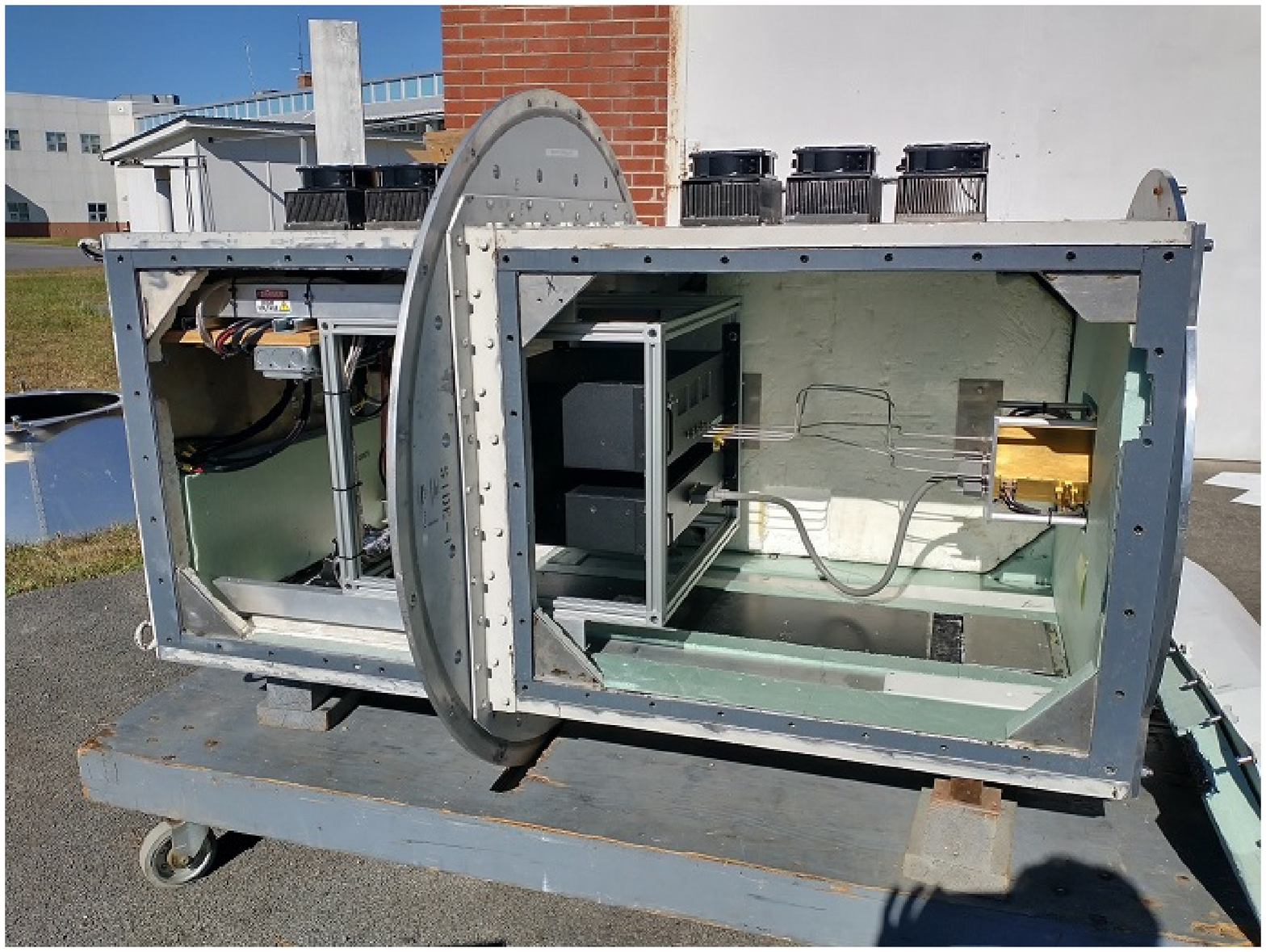}\\
\caption{Some photos of relevant instrumentation for the 310~MHz absolute map project.  The top photo shows the custom high edge-taper feed which will be mounted at the prime focus of the GBT, while the bottom photo shows the custom balanced correlation receiver mounted in a spare GBT prime focus receiver box.  The 310~MHz map will have an absolutely calibrated zero level and polarization information, which will be valuable for CMB, 21-cm, and other studies as discussed in \S \ref{GBTsec}.}
\label{SingalFig}
\end{figure}

These projects attest to the importance and impact of workshops such as the one that is the subject of this report.  The 310~MHz absolute map project was conceived and prioritized at the previous RSB workshop summarized in \citet{CP1}, and the anisotropy measurement at 140~MHz with LOFAR was conceived at a previous BAM meeting in 2018.

\begin{table*}[!htbp]
\scriptsize
\caption{All Relatively Recent Radio Maps Reporting an Absolute Zero-Level Calibration}
\label{tab1}
\vspace{-3mm}
\begin{center}
\begin{tabular}{cccl}
{\it Frequency (MHz)} & {\it Map} & {\it Instrument} & {\it Source of Absolute Zero-Level Calibration} \\
\hline
22 & \citet{Roger99} & Dipoles above  & Scaling relative to 408~MHz \citet{Haslam} \\ 
 & & reflecting screen & map at Zenith applied to other elevations\\
\hline
45 & \citet{Maeda99} & Phased array & Overlap with \citet{Alvarez97}, itself based \\ 
 & & Yagi dipoles & on small region overlap with unspecified pedigree \\
\hline
40,50,60,70,80 & \citet{DT18} & 240 dipole array & Flux calibrator sources \\ 
 & & with synthesized beam & and instrument gain modeling\\
\hline
408 & \citet{Haslam} & Jodrell Bank 75~m & Overlap with blocked-aperture, 7.5$^{\circ}$ resolution dipole \\
 & & clear aperture dish & measurement of \citet{PS62}\\
\hline
1420 & \citet{RR86} & Stockert 25~m & Small overlap with wide beam horn-based \\ 
 &  & blocked-aperture dish & measurement of \citet{HS66}\\
\hline
2300 & \citet{Tello07} & blocked-aperture dish & Total power radiometer  \\ 
 &  &  & calibrated with observations of moon \\
\hline
2326 & \citet{Jonas98} & HartRAO 26~m & Small overlap with horn-based south celestial  \\ 
 &  & blocked-aperture dish & pole measurement of \citet{Bers94}\\
\hline
\end{tabular}
\end{center}
\end{table*}

\subsubsection{310 MHz Absolute Map}\label{GBTsec}

The 310~MHz absolute map will be made by utilizing the unique features of the Green Bank telescope (GBT) along with custom instrumentation to enable an accurate absolutely calibrated zero-level.  The GBT is the world's largest clear-aperture telescope, allowing an observation of the radio sky without reflections and emissions off of supporting structures, is fully steerable to all azimuthal angles, allowing for the entire visible sky to be mapped and for scans which repeatedly pass through the north celestial pole (NCP), and is located in the National Radio Quiet Zone allowing for minimal radio frequency interference (RFI).  The custom instrumentation includes a unique, high edge-taper antenna feed which will be mounted at the prime focus of the GBT and will underilluminate the GBT dish, and a newly designed balanced correlation receiver, both visualized in Figure \ref{SingalFig} and discussed in detail in \S \ref{Bordenave}.

Maps of the diffuse radio emission are of the utmost importance in astronomy and astrophysics, including for CMB and 21-cm cosmology studies, as evidenced by the 408~MHz \citet{Haslam} map having over 1000 citations.  However it is the case that these maps did not have an absolute zero-level calibration as a primary goal, and such a calibration is typically derived, as is the case for the \citet{Haslam} map, from blocked-aperture observations with limited overlaps with previous, decades-old measurements made with low-resolution dipole antennas such as that of \citet{PS62}.  Dipoles have a beam pattern, and blocked apertures have reflections and emissions, which couple an uncertain amount of radiation from the bright Galactic plane, the ground, and other sources into measurements.  These absolute calibrations may also have depended on observations of standard flux calibrator sources and instrument gain modeling, which requires extrapolations over orders of magnitude in instrument response and the level of the RSB as an unknown offset to flux calibrators.  Table \ref{tab1} lists all radio frequency maps reporting an absolute zero-level calibration available in the literature from the past 40~years and how they determined their absolute zero-level.  It can be seen that all current knowledge of the actual level of diffuse astrophysical emission below 2~GHz ultimately derives from dipole-based and/or over 50-year-old low-resolution measurements.  

It can therefore be said that no large-area mapping of the diffuse radio emission at MHz frequencies with an absolute zero-level calibration as a primary goal has \emph{ever} been made.  Therefore the new 310~MHz map, which will also include full Stokes-parameters polarization information, will provide an an essential resource for understanding and constraining almost all Galactic and extragalactic phenomena that manifest in, or depend on the understanding of, diffuse radio emission, in addition to definitively measuring the absolute level of the RSB at MHz frequencies.

The project is scheduled to have an initial, overnight observing run on the GBT in Fall 2022 which will result in a porous, part-sky map.  The full map will require one night of observing in roughly every-other calendar month, and will require mounting and de-mounting of the custom hardware, and will progress subject to the availability of funds.  

\subsubsection{140 MHz Anisotropy Measurement}\label{AM}

In a recent work \citep{LF21} we presented the first targeted measurement of the anisotropy power spectrum of the RSB.  We did this measurement at 140~MHz where it is the overwhelmingly dominant photon background.  We determined the anisotropy power spectrum on scales ranging from 2$^{\circ}$ to 0.2\arcmin\ with LOFAR observations of two 18~deg$^2$ fields --- one centered on the Northern hemisphere coldest patch of radio sky where the Galactic contribution is smallest and one offset from that location by 15$^{\circ}$.  We found that the anisotropy power is higher than that attributable to the distribution of point sources above 100~$\upmu$Jy in flux.  This level of radio anisotropy power indicates that if it results from point sources, those sources are likely at low fluxes and incredibly numerous, and likely clustered in a specific manner.  This measurement and its implications are discussed in detail in \S \ref{Heston}.

\subsection{A 310 MHz Absolute Map --- David Bordenave}\label{Bordenave}

In order to make a new, modern absolutely calibrated zero-level map of the diffuse radio emission as discussed in \S \ref{GBTsec} we are employing several essential strategies.  These are:
\begin{enumerate}
    \item Utilizing the GBT which has a clear aperture, and thus is not subject to unknown emissions and reflections related to structures in the light path, and is fully steerable, allowing scans at constant azimuth to pass through the north celestial pole (NCP) every $\sim 15$~minutes to provide an unchanging reference point on the sky to verify the gains and receiver noise temperatures.
    
    \item A custom, high edge-taper feed which underilluminates the GBT dish, greatly reducing spillover pickup from the ground and any other structures beyond the edge of the dish.
    
    \item A custom balanced correlation receiver which will allow the gain scale to be calibrated absolutely and the receiver noise temperature to be known and near zero.
\end{enumerate}

A photograph of the custom feed is shown in the top panel of Figure \ref{SingalFig}.  It is constructed out of a frame and wire mesh to reduce weight and wind loading and assembles in six segments. The feed was developed with extensive modeling in CST and GRASP8 and its beam pattern has been measured on the GBO test range.    Its response is $\sim$15~dB at 39$^{\circ}$ off axis, which corresponds to the edge of the GBT dish, thus greatly reducing spillover emission pickup from the ground to around $\sim11$~K.  The residual spillover pickup can be estimated with tip scans of the GBT, resulting in a spillover uncertainty of just $\sim$2~K.

\begin{figure}
\includegraphics[width=3.4in]{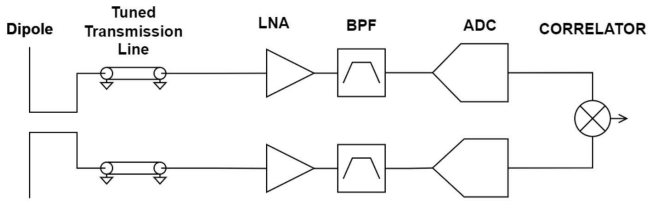}\\
\caption{Diagram of the balanced correlation receiver design.  `LNA' is low-noise amplifier, `BPF' is band pass filter, and `ADC' is analog-to-digital conversion.  There is one independent receiver chain as visualized here for each polarization. }
\label{BordenaveFig}
\end{figure}

A photograph of the receiver as constructed is shown in the bottom panel of Figure \ref{SingalFig}.  The housing seen on the right end contains the front end amplifier and calibration boards, and the black-paneled housing seen in the middle contains the digitization and control components.  These are seen mounted in a spare GBT prime focus receiver box, which will be installed on the GBT in place of the existing prime focus receiver box, while the custom feed will be attached to the front end of the receiver box.  

In order to achieve an accurate and precise absolute zero-level calibration, the receiver has a novel balanced correlation design to ensure gain stability and a known and low receiver noise temperature.  A receiver design often used for gain stability is the so-called pseudocorrelation receiver, where the output power of a dipole is divided into two isolated receiver chains after being referenced to ground with a balun.   Noise in the amplifiers and filters is uncorrelated between the resulting two channels but the sky power is correlated, allowing the statistical elimination of these noise fluctuations in a recombined signal \citep[e.g.][]{Wollak95}.  However in such a design any ohmic losses in the transmission line from the feed, balun, and common regions of the power divider are correlated and therefore add a noise temperature to the sky power measurement, the uncertainty of which is then a source of uncertainty in the absolute zero level.  In our balanced correlation receiver, rather than splitting the signal after the balun, the voltage signal on each arm of a given dipole is separately referenced to a common ground with its own transmission line and coaxial transition to eliminate this source of correlated noise.  A block diagram of the design is shown in the top panel of Figure \ref{BordenaveFig}.  

There is one independent receiver chain as visualized in the top panel of Figure \ref{BordenaveFig} for each polarization -- thus there are four chains of amplifier, band pass filter, and analog-to-digital conversion.  On the front end amplifier boards all channels can be switched to either their antenna arm or a 50~$\Omega$ resistive termination, providing a source of calibrated, uncorrelated noise to each channel.  In addition, common high and low level calibration noise sources are split and injected into all channels to produce correlated noise suitable for hot/cold Y-factor measurements of the receiver noise.  Analog-to-digital conversion takes place in a pair of Ettus Research B210 software-defined radio (SDR) modules, each processing the pair of channels associated with a given dipole.  Within the SDRs, the signals are divided into in-phase (``I'') and 90$^{\circ}$ out of phase (``Q'') components, mixed down, and digitized.  These signals are then fast Fourier transformed (FFTs) and correlated in back end software to give the measured autocorrelated power for each channel and cross-correlated power for all six pairs of channels in spectral bands of $\sim$1~MHz in real time and $\sim$100~kHz upon further processing over the 20~MHz band.  Residual RFI can be filtered further with kurtosis of the spectral signal given these narrow bands.

This will allow the absolute zero-level calibration to be achieved and maintained as follows: i) Each channel will have its gain measured by recording its output autocorrelation power when its input is terminated in physical loads at 77~K and room temperature.  ii) A measurement of the autocorrelation power when viewing the high and low internal calibration loads and application of the measured channel gain determines the true precise effective emission temperature of these loads for the channel.  iii) According to basic receiver theory the gain for the cross-correlation of two channels is the geometric mean of the gains of the two channels.  iv) The receiver noise temperature should be zero or very close to it for the cross-correlations because all noise for any two channels is uncorrelated in the balanced correlation design, and this can be verified by measuring the output cross-correlation power for terminating in the physical loads.  v) With the gain and receiver noise temperature known for all cross-correlations, all Stokes parameters can be determined, with the absolute intensity $I$ being the sum of the two cross-correlations across channels of the same polarization, and the parameters describing the polarization, $Q,U,V$, being sums and differences of the four cross-correlations across channels of opposite polarization.
\begin{eqnarray}
    I = X_1 X_2^* + Y_1 Y_2^* \\
    \nonumber Q = X_1 X_2^* - Y_1 Y_2^* \\
    \nonumber U = X_1 Y_2^* - Y_1 X_2^* \\
    \nonumber V = i (-X_1 Y_2^* - Y_1 X_2^*)
\end{eqnarray}
where $X$ and $Y$ represent the two linear polarizations.  vi) The cross-correlation gains and receiver noise temperatures (which should be zero) can be constantly verified in-situ during observing by switching in the internal noise sources, and because the whole system will view the unchanging NCP every 15~minutes.

The performance of the balanced correlation receiver has been extensively modeled down to the individual circuit element level in Keysight ADS with the model parameters determined with extensive vector network analyzer measurements of the S-parameters of the elements.  The receiver noise temperatures and their uncertainties are low enough that the total uncertainty in the absolute zero level due to the receiver is $\sim$4~K, which adds in quadrature to that due to spillover pickup for a total zero-level uncertainty of 5~K, much less than the sky temperatures at 310~MHz.

\subsection{Anisotropy of the RSB at 140 MHz --- Sean Heston}\label{Heston}

\begin{figure}[!htbp]
\includegraphics[width=3.2in]{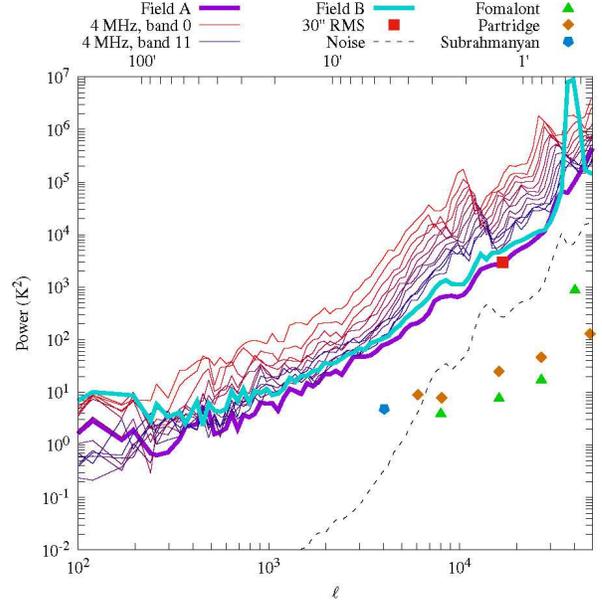}\\
\caption{Measured anisotropy power spectrum of the radio sky centered at 140~MHz with an RMS noise of 720 $\mu$Jy. Shown are curves for the full bandwidth of the coldest patch field (field A) and the secondary field (field B), as well as for 12 four MHz wide sub-bands of field A. The anisotropy in field A deduced by considering the average noise per beam in the image with the synthesized beam tapered to 30\arcsec\ FWHM is also shown and agrees at the relevant angular scale. We also show comparison levels inferred by the noise per beam at 8.7~GHz, 8.4~GHz, and 4.9~GHz in different fields as calculated by \citet{Holder14} and scaled here to 140~MHz assuming a synchrotron power law of -2.6 in radiometric temperature units. The amount of angular power is $\sim$1.4 times higher for field B compared to field A (in K$^2$ units) across a range of angular scales. All angular powers are expressed here in the $\left({\Delta T} \right)_\ell^2$ normalization. Reproduced from \citet{LF21}}
\label{HestonFig}
\end{figure}

As mentioned in \S \ref{AM}, we performed the first targeted measurement of the power spectrum of anisotropies of the RSB at MHz frequencies, where it is the dominant photon background (see top panel of Fig.~\ref{CRB}). This area of RSB research is relatively unexplored. Previous studies of temperature power spectra for different frequencies have helped to confirm the source populations for the cosmic infrared \citep[e.g.,][]{Planck11, George15} and gamma-ray \citep[e.g.,][]{Broderick14} backgrounds, and have been a critically important part of CMB science so far \citep[e.g.,][]{Planck:2018vyg}. 

We measured the anisotropy power spectrum of the RSB with two observation fields, each 18~deg$^2$, taken by LOFAR at 140 MHz. Our primary field (Field A) was centered on the Galactic Northern Hemisphere ``coldest patch'' \citep{Kogut11}: $9^h\,38^m\,41^s$ +$30^{\circ}49'12''$, $l=$196.0$^{\circ}$ $b$=48.0$^{\circ}$, which is the area of lowest measured diffuse emission absolute temperature and thus where the integrated line-of-sight contribution through the Galactic component is minimal. LOFAR allows for a simultaneous observation of a separate field offset by 15$^\circ$ in an adjacent 48 MHz wide band, so we selected a location closer to the Northern Galactic Pole from the coldest patch (Field B) $10h\,25m\,00s$ +$30^{\circ}00'00''$ ($l$=199.0$^{\circ}$ $b$=57.9$^{\circ}$). Field B should have more, but nearly minimal, total Galactic contribution when compared to Field A.

We performed direction-independent calibration on the two measurement fields, using a manual calibration approach for the coldest patch field (Field A) and an automated calibration for the offset field (Field B). The manual and automated calibrations have similar results, which is why Field B was calibrated using the automated approach. We then extracted the angular power spectra of the fields using the calibrated images. The details of the calibration and power spectra extraction processes are outlined in \citet{LF21}.

The results for the power spectra extraction process are shown as the thick purple (Field A) and cyan (Field B) lines in Fig.~\ref{HestonFig}. Also shown are twelve 4 MHz wide sub-bands of Field A (thin lines). The separation of these sub-bands comes from the spectral dependence of synchrotron radiation, causing the lowest frequency band to have ${\sim}17$ times more power than the highest frequency sub-band. The full bandwidth (Field A) has lower angular power due to a more complete $u{-}v$ coverage. The red square is the 30'' scaled RMS noise of Field A calculated by the procedure described in \citet{Holder14}, which agrees at the relevant angular scale. Older GHz scale measurements are also shown as triangles, diamonds, and a pentagon, again calculated using \citet{Holder14}. Finally, we show the estimated contribution from system noise as the dashed line. We see that our measured power is much higher than what is suggested by the previous GHz scale measurements. Our measurements are also not dominated by system noise, as seen by the large spacing between the measured field lines and the dashed noise line. 

Field B's angular power is larger by a factor of ${\sim}1.4$ in the $(\Delta T)^2$ normalization (therefore ${\sim}1.2$ for $\Delta T)$ than the angular power Field A. This factor is also the square of the ratio of the average absolute brightness of the two fields in radiometric temperature (K) units calculated from the map of \citet{Haslam}. The differences of observed absolute brightness between the fields should only come from differences in lines of sight through the Galactic diffuse component. This is a strong indication that the proportion of angular power, in $(\Delta t)$ units, due to Galactic structure traces the proportion of absolute brightness due to that structure, for lines of sight with minimal Galactic structure and probably for lines of sight far away from the Galactic plane. We believe that this suggests that the contribution from Galactic is sub-dominant as the \textit{normalized} angular power $\left(\frac{\Delta T}{T} \right)$ is the same for both fields and the Galactic structure is what varies spatially between our two fields.

In order to account for possibly unremoved point sources below detection threshold in our observation fields, we created a Monte Carlo catalog of simulated sources from 100$\mu$Jy to 40 mJy following the flux distribution of \citet{Franzen16}. We placed these sources both randomly in RA and Dec as well as placed using simple sinusoidal clustering on scales of 1' and 10'. We then imaged the point source files and ran them through the power spectrum pipeline. The results for this analysis are shown in Fig.~5 of \citet{LF21}. We found that the measured anisotropy power cannot be attributed to potentially unremoved point sources that follow the Franzen flux distribution.

We then decided to test much dimmer and more numerous point sources, specifically a model from \citet{Condon12} with the least number of sources, which is meant to reproduce the measured radio surface brightness of the sky. We again modeled these point source distributions with random positions as well as sinusoidally clustered, but only on scales of 1'. The resulting power spectra of these source populations are also shown in Fig.~5 of \citet{LF21}. We found that neither model, with and without clustering, produced enough angular power. However, we saw that the clustered model had increased angular power on all angular scales smaller than the clustering scale. Therefore, we are investigating whether a model of very faint but extremely numerous point sources, with the right clustering on multiple angular scales, can reproduce the measured anisotropy power.

\subsection{Source Populations in the Extragalactic Radio Sky --- Catherine Hale}\label{Hale}

Our knowledge of the total radio background level that is specifically contributed by known extragalactic source classes is being transformed by recent surveys, which are allowing us to push deeper to understand whether the contribution of faint radio sources can be reconciled with measurements of the sky background level as discussed in \S \ref{overview}. Works such as \citet{Vernstrom14}, \citet{Murphy18}, \citet{Hardcastle21}, \citet{Matthews21}, and \citet{Hale22} have all studied contributions of extragalactic radio sources to the excess sky background temperature (between 144~MHz -- 3~GHz) but all find total temperature contributions from extragalactic sources a factor of $\sim$4 smaller than the RSB level discussed in \S \ref{intro}. Below the nominal 5$\sigma$ detection limit, extrapolations of the source counts using stacking \citep[see e.g.][]{Zwart15} and $P(D)$ analysis \citep[see e.g.][]{Vernstrom14, Matthews21} from extragalactic radio images are unable to detect an extremely numerous faint population of sources that would reconcile with the measured RSB level. Recent work by \citet{Matthews21} has consider the contribution of both AGN and star-forming galaxies (SFGs) to the extragalactic sky background temperature combining $P(D)$ analysis for the source counts and using evolving local luminosity functions at $z=0$ for AGN and SFGs. They find a limiting total background temperature contribution of $\sim$110 mK at 1.4~GHz even down to ~10 nJy. This work therefore suggests that the known extragalactic contributions from AGN and SFGs cannot account for the measured level of the RSB.

As we move to the future of radio surveys, directly detecting faint radio populations (and being able to probe below 5$\sigma$) will rely on surveys from telescopes such as the Australian Square Kilometre Array Pathfinder \citep[ASKAP --][]{Johnston07,Hotan14,Hotan21}, the Meer Karoo Array Telescope \citep[MeerKAT --][]{Jonas09,Booth09,Jonas16}, LOFAR \citep{VH13} and eventually the next generation Very Large Array \citep[ngVLA --][]{Murphy18b} and the Square Kilometre Array Observatory \citep[SKAO --e.g.][]{Wilman08}. These facilities will all push observations to unprecedented sensitivities. However, with this increased depth also comes challenges of source confusion which is only possible to be overcome by high angular resolution. Recently, at low frequencies (<200~MHz), LOFAR has demonstrated that they are able to overcome such resolution issues, making use of their array stations spread across Europe \citep[see e.g.][]{Morabito22,Sweijen22}. At higher frequencies ($\sim$1 GHz), sub-arcsecond imaging has been observed using telescopes such as e-MERLIN \citep[e.g.][]{Muxlow20} and the Very Long Baseline Array \citep[VLBA, see e.g.][]{HerreraRuiz17} and at higher frequencies still, sub-arcsecond resolution is already possible with surveys such as the VLA for frequencies at S-band and above \citep[see e.g.][]{Smolcic17}. However, the combination of depth, sensitivity and resolution that allows us to determine the contribution of faint sources to the extragalactic radio background and minimize the effects of cosmic variance will rely on LOFAR, ngVLA and SKAO observations. With these we will be able to probe to sub-$\mu$Jy depths and consider whether an even fainter population of extragalactic sources can account for the level of the RSB.

\subsection{The LWA1 Sky Survey --- Chris DiLullo}\label{Dilullo}

Low frequency measurements of the sky below 200~MHz are important for determining the nature of the radio synchrotron background.  Combined with higher frequency observations, they can determine the spectral index of the background and help constrain if there is evidence of a spectral break in the power law. They are also  important for modern 21 cm cosmology experiments aiming to detect the redshfited 21 cm signal from neutral hydrogen present during Cosmic Dawn and the Epoch of Reionization as they map the Galactic foregrounds which have to be removed to detect the cosmological signal.

The LWA1 Low Frequency Sky Survey \citep[LLFSS;][]{Dowell17} offers some of the only zero-level absolutely calibrated maps of the sky below 100~MHz. The first station of the Long Wavelength Array, LWA1, is an array consisting of 256 dipoles within a 100~m~x~110~m area along with five outrigger antennas \citep{Taylor12}. It offers three data collecting modes: Transient Buffer Narrow (TBN), Transient Buffer Wide (TBW), and a beamformer mode. The TBN and TBW modes are raw voltage  modes which record the raw voltages from the antennas either continuously for a narrow bandwidth or for a short duration for the entire 80 MHz of bandwidth offered by the array, respectively. The survey was carried out by collecting  TBW data between 10 -- 88~MHz for a duration of 61~ms every 15~minutes over two days. These data were then correlated and imaged using the LWA Software Library \citep[LSL;][]{Dowell12}. 

The survey's total power calibration is derived from custom front end electronics boards which were designed for the Large Aperture Experiment to Detect the Dark Ages \citep[LEDA;][]{Price18}. These custom radiometers  are connected to the outrigger antennas and provide a means to measure antenna temperature via a three-state switching technique that is commonly used in other 21 cm experiments. LEDA data was used to provide a scaling between observed power and temperature for the survey data.

The LLFSS provides absolutely calibrated maps of the sky at nine frequencies ranging from 35 -- 80~MHz. These data have been used to estimate the temperature of the extragalactic radio background \citep{DT18}. In that  work, the authors model and remove the Galactic contribution to the LLFSS data and find that the remaining extragalactic contribution obeys an expected power law and is consistent with the ARCADE~2 results \citep{Fixsen11, Singal11}. A summary of the results can be seen in the top panel of Figure \ref{CRB}.  They also note that the extragalactic radio background, when considered with the results of ARCADE~2, shows no sign of a spectral break or turnover.  However, they note that the results are highly dependent on how the Galactic foreground is removed and also the underlying calibration of the LLFSS.

Current efforts to improve the LLFSS have focused on directly measuring the impedance mismatch between the LWA antenna and the front end electronics. The impedance mismatch correction in the current LLFSS is based on simulations  \citep{Hicks12} and is a key step in setting the flux scale for the entire survey.  Therefore, accurate measurements are necessary to improve the calibration of the 
survey. Preliminary measurements have been made and \textit{in situ} measurements at the telescope are being planned for the near future. A new survey is also underway  which will offer more data with which to build the sky maps and possibly increased  frequency coverage.

\subsection{C-BASS: An All-Sky Survey of Galactic Emission at 5~GHz in Intensity and Polarization --- Stuart Harper}\label{Harper}

The study of Galactic synchrotron emission in the optically thin regime has been dominated over the last few decades by the full-sky map of the Galaxy at 408\,MHz \citep{Haslam}. The 408\,MHz map has been critical to the success of recent CMB missions \citep[e.g.][]{Planck20}, and to our understanding of Galactic synchrotron emission produced by cosmic ray leptons propagating through the Galactic magnetic field \citep{rybicki/lightman:1979}. However, the 408\,MHz is limited to total intensity only. Measurements of the polarized Galactic emission will be required to study features of the Galactic magnetic field, and will be necessary for future CMB missions such as Simons Observatory \citep{Ade2019} and LiteBIRD \citep{Hazumi2020} to detect the polarized B-mode emission produced by primordial gravitational waves. To date there are very few maps of Galactic polarized emission. The 1.4\,GHz Dominion Radio Astronomy Obesrvatory (DRAO)/Villa-Elisa all-sky map \citep{Wolleben2006,Testori2008} has been shown to have many systematic errors \citep{Weiland2022}, while the 2.3\,GHz S-band Polarization All Sky Survey (S-PASS) map \citep{Carretti2019} only covers the Southern sky. Also, at frequencies below 5\,GHz Faraday rotation---polarization angle rotation of radiation traversing a magneto-ionic plasma---starts becoming a serious issue even at high latitudes. Simple corrections for Faraday rotation using data from extragalactic polarized sources \citep{Hutschenreuter2022} is non-trivial since the distances to many high latitude structures are uncertain, and likely have multiple Faraday screens along each line-of-sight.

The C-Band All-Sky Survey (C-BASS) project will produce an all-sky map of Galactic synchrotron emission at 5\,GHz with a resolution of 1\,degree full-width half-maximum in both total intensity and polarization. The C-BASS project is a collaboration between the University of Manchester and Oxford in the UK, Caltech in the US, and University of Kwazulu-Natal and the South African Radio Astronomy Observatory in South Africa. The project is a combination of a Northern survey based in the Owens Valley Radio Observatory, observations were taken between 2012 and 2015, and the final data processing is expected to be finished later in 2022. The Southern survey is based in the Karoo national park in South Africa, with observations currently ongoing. 

The Northern C-BASS instrument is a cryogenically cooled dual circularly polarized radiometer that can simultaneously measure Stokes I, Q, and U \citep{Jones2018}. The bandpass spans 4.5--5.5\,GHz but a notch filter is used to suppress the central 0.5\,GHz due to radio frequency interference (RFI). For the C-BASS South instrument the receiver design will be updated to include a 128 channel spectrometer that will allow for the accurate excision of RFI and the measurement of in-band Faraday rotation.

The C-BASS North telescope is a 6.1\,m dish with a Gregorian design while the C-BASS South has a 7.3\,m dish and a Cassegrain design. The C-BASS South primary is highly under-illuminated in order to match the main beam to that of the norther telescope. The C-BASS optics were designed to ensure a circularly symmetric beam pattern by having the on-axis secondary reflector supported by a low-loss dielectric foam cone instead of support struts. Far sidelobe contamination was minimized by surrounding the primary with a radio-absorbing baffle \citep{Holler2013}. 

\begin{figure}
    \centering
    \includegraphics[width=0.48\textwidth]{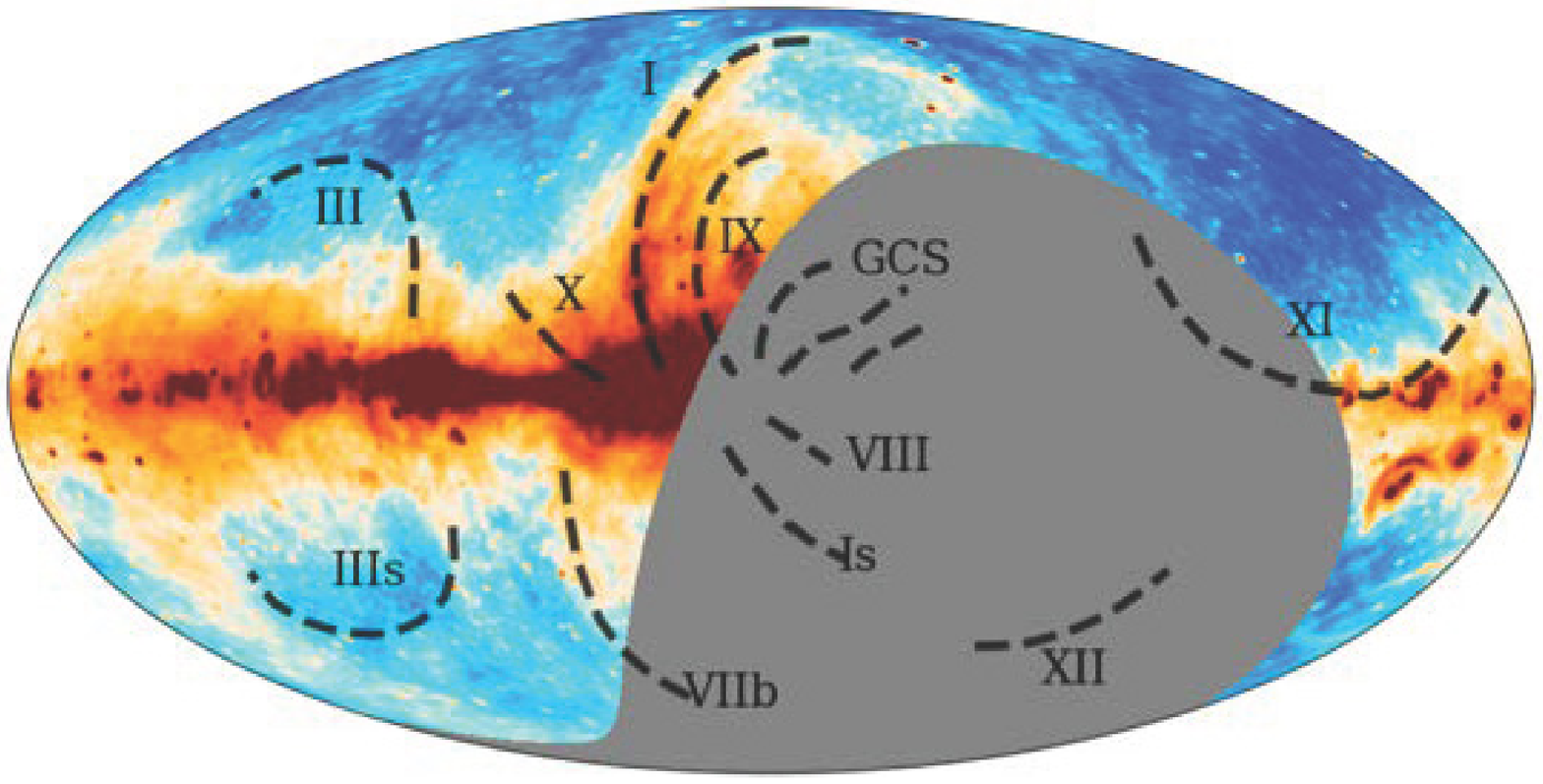}
        \includegraphics[width=0.48\textwidth]{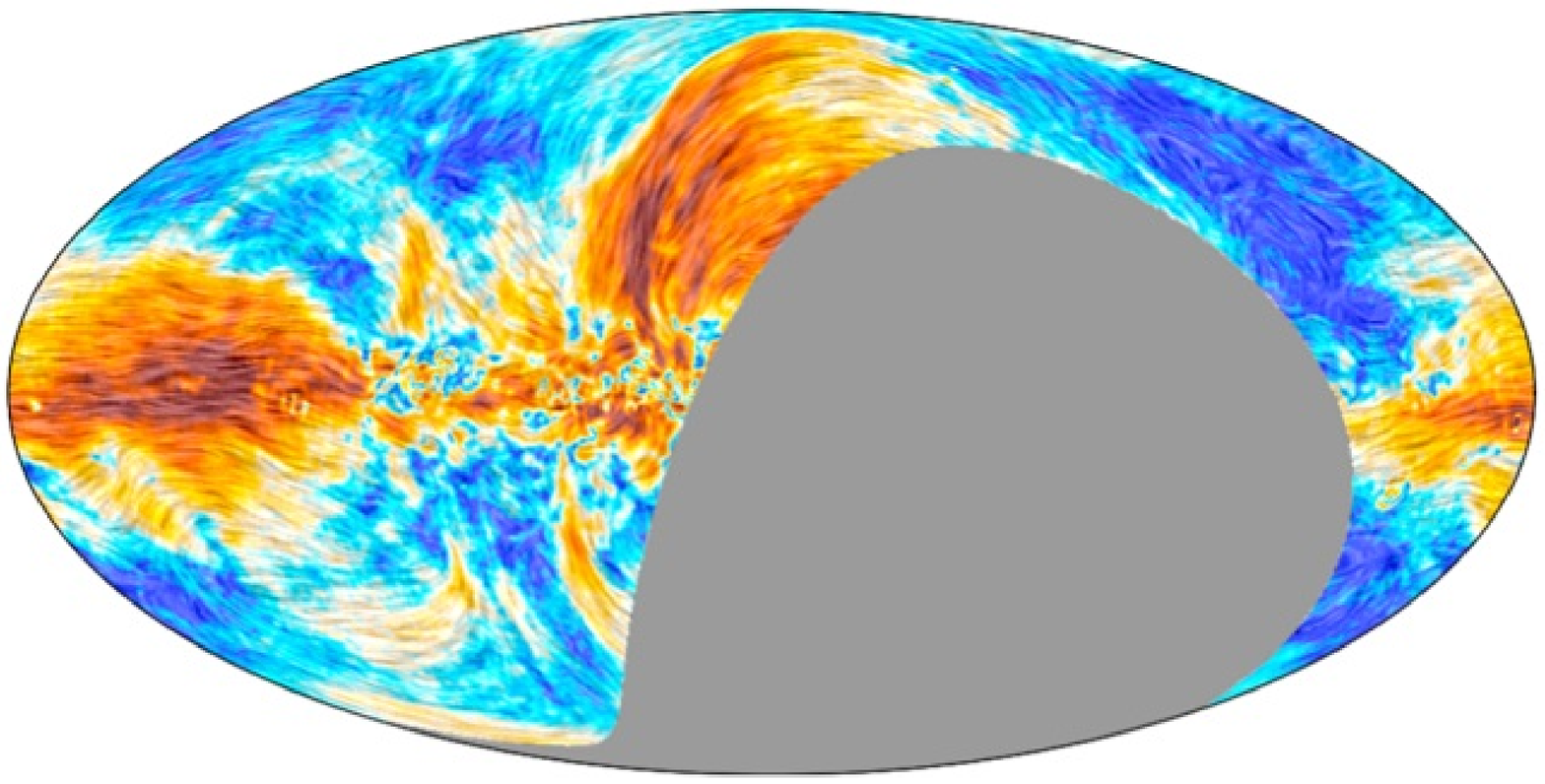}
    \caption{C-BASS North maps of total intensity (\textit{top}) and polarized intensity (\textit{bottom}) in Galactic coordinates using a mollweide projection. The total intensity map color scale has been saturated along the Galactic plane and includes the locations of well-known radio loops \citep{Vidal2015}. The polarized intensity map has the projected B-field direction overlaid using line-integral convolution.}
    \label{Harper:fig:cbass}
\end{figure}

The top-panel of figure~\ref{Harper:fig:cbass} shows the C-BASS North total intensity map, while the bottom-panel shows the polarized intensity map with the polarization vectors overlaid using the \texttt{Healpy} line-integral convolution (LIC) routine \citep{Gorski2005}. In total intensity the location of known radio loops \citep{Vidal2015} have been overlaid. The polarized intensity map has a noise level of 0.1--0.2\,mK r.m.s. per deg$^2$, which will allow for the constraint of polarized synchrotron down to a level of $\sim0.75$\,$\mu$K-arcmin$^2$ at 100\,GHz. The signal-to-noise of the C-BASS polarization data is greater than 5 for more than 95\,per\,cent of the sky at 1\,degree resolution, with Faraday rotation angles less than a 5\,degrees for regions away from the Galactic plane---for lower frequency surveys, like S-PASS, this value is typically a factor of 3--4 times larger.

The final C-BASS map will be the highest signal-to-noise and the most robust template of Galactic synchrotron emission for studies of the CMB and Galactic astrophysics for the foreseeable future.

\subsection{C-BASS: Polarization in the Northern Hemisphere: Fractional Polarization and Constraints on Field Tangling --- Patrick Leahy}\label{Leahy}

Synchrotron polarization is a powerful diagnostic of the structure of the Galactic magnetic field. Its intrinsic polarization is $\approx75$\% (for spectral index $\alpha \approx -1$), but, as observed, this is reduced by averaging of different field orientations along the line of sight and across the beam, as well as by differential Faraday rotation at long wavelengths. Synchrotron polarization therefore gives two independent measures of the tangling of the field: the observed pattern of the polarization angle $\chi$ on the sky, and its fractional polarization, $m$. Any extragalactic radio background is expected to have negligible polarization unless observed at a very high resolution that can resolve it into individual sources, since polarization orientation should not be correlated over cosmological
distances.

As described in \S \ref{Harper}, C-BASS provides our best view of the Galactic polarized emission, with minimal Faraday distortion and far higher signal-to-noise than {\it WMAP} and {\it Planck}. In particular, in those missions synchrotron cannot be accurately separated from other emission processes, leaving the fractional polarization uncertain by factors of several \citep[e.g.][]{Vidal2015}.

C-BASS does not measure the overall zero level, and so we have set this using the ARCADE~2 maps at 3, 8 and 10\,GHz \citep{Fixsen11}: after subtracting the CMB monopole and dipole, we interpolated to the C-BASS frequency of 4.76\,GHz assuming a power-law spectrum in each pixel, and fitted the result directly to the C-BASS map, convolved to ARCADE resolution and also with the CMB dipole subtracted, in the same set of pixels. We estimate  about 5\,mK uncertainty, limited by residual systematics in ARCADE.

\begin{figure}
\includegraphics[width=3.0in]{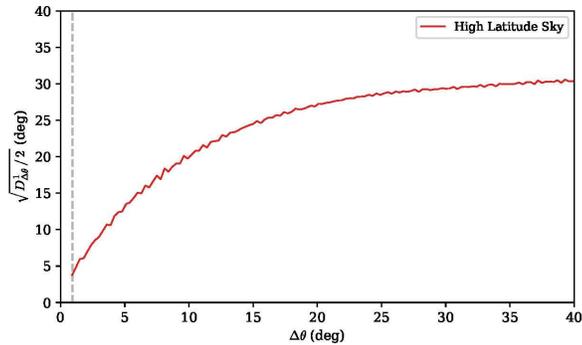}
\caption{$\sqrt{D/2}$ vs. separation $\Delta\theta$ for the high-latitude C-BASS polarization angles, where $D$ is the structure function, so $\sqrt{D/2}$ would equal the standard deviation, on scales large enough that pixel pairs are uncorrelated.}
\label{LeahyFig}
\end{figure}

Apart from a narrow region along the Galactic plane where Faraday depolarization is still significant, C-BASS shows a field pattern ordered on scales much larger than our 1\degr\ beam
(Fig.~\ref{Harper:fig:cbass}). This applies not only in the prominent discrete `loops' and ``spurs,'' but also in high-latitude regions well away from these structures, where lines of sight presumably sample typical paths through the thick synchrotron-emitting disk. To quantify this, we measured the position angle structure function, $D(\Delta\theta) = \langle(\chi(\hat{\bf n})-\chi(\hat{\bf m}))^2\rangle$, where the average is over all pairs of directions $\hat{\bf n}$ and  $\hat{\bf m}$ separated by angle $\Delta\theta$. We used parallel transport to ensure this function is coordinate independent, and polarization angle differences are folded into $|\Delta \chi| \le 90\degr$. Random polarization angles would give $\sqrt{D/2} = 37\degr$. The results are plotted
in Fig.~\ref{LeahyFig}.
We restricted our analysis to the northern high-latitude sky, $b > 30^\circ$, excluding regions affected by the high-latitude Loops I and III. $\sqrt{D/2}$ flattens off at about $\Delta \theta_0 = 15\degr$, although there is residual large-scale order since the value stays significantly below 37\degr\ out to beyond $\Delta\theta = 40\degr$. For a simple model in which the field is coherent over scales $d$ on a path-length $L$, we expect $N=L/d$ cells on a line of sight, and the structure function will flatten at $\sin(\Delta\theta_0/2) \approx 1/N$; so our result suggest $N\approx 7.7$. This model gives a random walk in polarization space $(Q,U)$ with $N$ steps, hence reducing the polarization fraction by $\sqrt{N}$, so predicts $m = 75\%/\sqrt{7.7} \approx 27\%$. This is highly inconsistent with the observations, which have $\langle m\rangle = 3.3$\% in the same region, and hardly any pixels with $m > 10\%$. The low observed $m$ in this `empty' region of sky also contrasts with much higher values in discrete features like Loop I (the North Polar Spur), where the raw polarization reaches up to $\approx 30$\% without any correction for an underlying weakly-polarized background.

At first sight, we might invoke an unpolarized isotropic background to resolve the paradox. But this does not work (although see \S \ref{Kogut}): the ARCADE model of \citet{Seiffert11} implies a contribution of 16.4\,mK at our frequency, which reduces the average Stokes $I$ brightness in our region by less than a factor of two, rather than the factor of eight required. Nor are we convinced by the original arguments of \citet{Kogut11} for such an extragalactic background: the misfit to a slab model ($\csc|b|$ law)
or tracers of the thin disk, such as [C{\sc ii}], is anisotropic, so at least partly local: the minimum synchrotron emission in both hemispheres is at intermediate latitudes, and there are large variations between galactic quadrants even excluding the loops. We see no reason to invoke a separate cosmological excess as well.

A second resolution would invoke multi scale tangling in the magnetic field, so that the large-scale order visible in the polarization angles is superimposed on a much finer-scale tangling unresolved by C-BASS. Such a single-scale model is doubtless oversimplified, but a power-law spectrum of field fluctuations cannot resolve the problem: only a very steep angular power spectrum can produce the large-scale order observed, and then the small-scale fluctuations are too weak to cause depolarization. \citet{Leclercq2017} measured the angular power spectrum of the diffuse polarized emission on scales down to 3\farcm 4 in the Arecibo G-Alfa Continuum Survey \citep[GALFACTS --][]{Taylor2010}, and found that $C_\ell^{E,B}$ mostly declined faster than $\ell^{-2.6}$ in the set of 15\degr$\times$15\degr\ region considered,  which is too steep for the fine-scale fluctuations to cause substantial depolarization.

A third resolution would be a fortuitous cancellation of polarization along the line of sight, for instance if the thick-disk (or halo) field was nearly orthogonal to that in the thin disk. 

Since none of these resolutions are particularly satisfactory, the low fractional polarization of the high-latitude synchrotron emission remains a major puzzle, and merits further analysis including more realistic modelling.

\subsection{A Cross-correlation Analysis of CMB Lensing and Radio Galaxy Maps --- Giulia Piccirilli}\label{Piccirilli}

Besides the large amplitudes of the RSB discussed in \S \ref{overview} and that of the dipole in the spatial distribution of the radio sources  (see \citet{Peebles22} and references therein), there is yet one more anomaly that characterizes the radio sky: the high amplitude of the two-point autocorrelation function of the radio sources in the TGSS-ADR1 catalog \citep[TGSS -- ][]{Intema17} at large angular scales \citep{Dolfi19}. Whether this is a genuine feature or an artifact due to unidentified systematic errors in the data analysis, it is a question that we have addressed by cross correlating the angular position of the TGSS sources (since only a small fraction of them have measured redshifts) with the angular map of an unbiased tracer of the underlying mass density field. For the latter, we considered the lensing map of the CMB \citep{Planck20}.
The motivations for studying this cross correlation are several. Firstly, the two maps are expected to be prone to different systematic errors. Even when they are not properly identified and corrected for, these errors are not supposed to correlate with each other and therefore should not contribute to the cross-correlation statistics. In addition, correlating the angular position of some biased mass tracer of the mass field, like the radio sources, characterized by a redshift-dependent bias $b(z)$ and a redshift distribution $N(z)$, with that of unbiased tracers allows us to break the degeneracy of these two functions. This is clear from the expression of the cross angular spectrum below:
\begin{eqnarray}
\label{eq:cross_spectrum_theory}
\ = \frac{2}{\pi}\int^{\infty}_0 dz W^g(z)\int^{\infty}_0 dz' W^{\kappa}(z') \times 
\\ \nonumber \int^{\infty}_0 dk k^2 P(k;z;z')j_{\ell}[k \chi(z)]j_{\ell}[k \chi(z')].
\end{eqnarray}

In Equation~\ref{eq:cross_spectrum_theory}, the window function $W^g(z)$ of the biased tracers depends on the product $b(z)\times N(z)$, whereas the window function of the lensing signal $W^{\kappa}(z)$ does not. Therefore, combining the cross spectrum with the auto spectrum of the tracers (that is proportional to [$W^g(z)]^2$) and with that of the lensing signal (that depends on $[W^{\kappa}(z)]^2$) can potentially break this degeneracy. As a result, the cross correlation analysis is expected to be less biased and, in combination with the autocorrelation, it is able to provide information on clustering properties and on the nature of the radio sources.
\begin{figure}
    \centering
        \includegraphics[width=\linewidth]{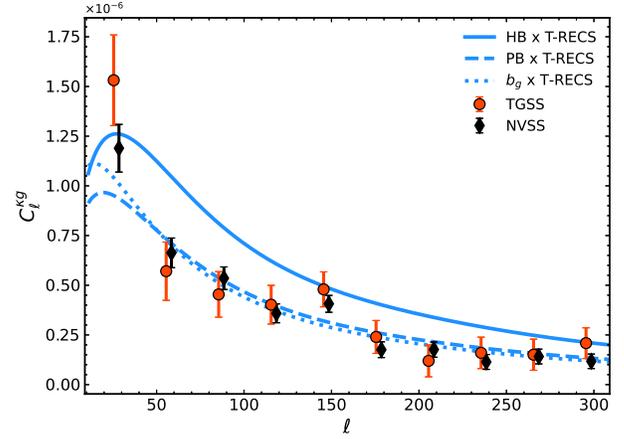}
    \caption{Measured cross spectrum $\kappa g$ of TGSS -- CMB lensing (red dots) and NVSS -- CMB lensing (black diamonds).  Error bars represent the $1\sigma$ uncertainties obtained assuming Gaussian statistics. Blue curves represent model predictions obtained using different prescriptions for the bias, $b(z)$, while the T-RECS $N(z)$ distribution of \cite{Bonaldi18} is used (see \citet{Piccirilli22} for a detailed description).}
    \label{PiccirilliFig}
\end{figure}
From our cross-correlation analysis we obtained three main results:
\begin{itemize}
\item First of all we detected the TGSS -- CMB convergence cross-correlation signal with $12\sigma$ significance. As  shown in Figure~\ref{PiccirilliFig}, the measured cross spectrum is in good agreement with the one obtained using the radio sources of the NVSS catalog \citep[NVSS -- ][]{Condon98}, whose auto-spectrum does not show a similar excess. We then conclude that the power excess originally detected in the TGSS auto-spectrum  on large angular scale probably originates from unidentified systematic observational effects.

\item After having verified the genuine nature of the two cross-correlation signals, we tried to fit the measured cross-spectra with theoretical models for $b(z)\times N(z)$ taken from the literature. 
To do so, we performed two $\chi^2$  tests: the first one uses the cross-spectrum only, while the second one uses both the cross- and auto-spectra. 
For the $b(z)$ model of both TGSS and NVSS sources we considered two cases, both based on the Halo Model \citep[HM -- ][]{Cooray02}: i) the Halo Bias model of \citep[HB -- ][]{Ferramacho14}, in which different types of radio sources are assigned to halos of different masses and biases; ii) the Parametric Bias model (PB) proposed by \citep{NT15} to fit the angular spectrum of the NVSS sources.\\
For both TGSS and NVSS catalogs, we found that models that can fit the large angular scales
overpredict the power on smaller scales and are ruled out  (e.g. $\chi^2 = 4.88$, HB model in Figure~\ref{PiccirilliFig}) while the other ones perform similarly well in describing the behavior of the estimated spectra on all scales but the larges ones (e.g. $\chi^2= 1.59$, PB model  in Figure~\ref{PiccirilliFig}). These results are robust against the choice of the $N(z)$ model and to the potential systematic errors that may affect the radio catalogs.
\item  As the models proposed in literature to match the NVSS and TGSS auto-spectra do not provide a good fit to the cross-spectra, we tried to constrain the $b(z)$ model from data keeping the $N(z)$ and cosmological model fixed. We did not leave $N(z)$ free to vary since, as we have seen,  our results are quite insensitive to the choice of different redshift distribution model.
Following \citet{Alonso21} and given the limited number of data points in our analysis, we considered two simple bias models which depend upon one single parameter, the effective linear bias $b_g$. Focusing on the joint analysis of the NVSS catalog, we find that the non-evolving $b_g$ model (shown with a dotted line in Figure~\ref{PiccirilliFig}) fits both the auto- and cross-spectrum on large angular scales better than the  one that evolves with the redshift. The reason for this is that the constant bias model gives more statistical weights to the nearby, low redshift sources that dominate the cross-correlation signal at low multipoles,
\end{itemize}
Our results indicate that, even after reducing the contribution of spurious signals through the cross-correlation technique, the clustering amplitude of the radio sources on angular scales of ~$10^{\circ}$ remains large with respect to $\Lambda$CDM prediction for a population of radio objects with an $N(z)$ consistent with their observed luminosity function and with the expectation of the halo model bias. The excess is not large but is systematically detected in all models explored. The excess can, at lest in part, accounted for by advocating a large, constant bias factor $b_g$ with magnitude comparable with that of the bright QSOs at high redshift which, however, is difficult to reconcile with the presence of the mildly biased star forming galaxies that dominates the population of the radio sources at low redshifts. 
Alternative models of a decreasing bias as a function of redshifts proposed by \citep{Negrello06} did not improve the quality of the fit. Our results therefore hint at a large-scale clustering excess of the radio sources in the 100~MHz-1~GHz band, but are not conclusive with respect to its interpretation. For that we will have to wait for next generation wide surveys of a much larger number of sources like SKA precursors or, on the shorter term, for complementary analyses in other radio bands like the one that is being carried out by the LOFAR team.

\subsection{The Radio SZ Effect --- Gil Holder}\label{Holder}

The Sunyaev-Zel'dovich (SZ) effect \citep[e.g.][]{SZ80} arises due to the inverse-Compton upscattering of photons by energetic electrons, causing a distortion to the base photon spectrum.  It was originally, and has commonly been, understood in the context of the CMB, where the hot electrons of galaxy clusters distort the CMB blackbody in a detectable way when viewed in the direction of a cluster, with an increment in the observed surface brightness and radiometric temperature at higher frequencies and a decrement at lower frequencies.  The CMB SZ effect has been a major focus of CMB science and the study of clusters for cosmology and other purposes \citep[e.g.][]{Mr19}.   
In \citet{HC22} we noted that the RSB, if it is extragalactic, would act as an ambient photon field for clusters in the same manner as the CMB, and therefore one would expect a ``radio SZ'' effect to distort the power-law spectrum of the RSB.  Since the base RSB signal is a continuous power law, an SZ signal for the radio alone would be an increment at all observable frequencies.  However in that work we showed that because the CMB is also an appreciable contributor to the total surface brightness at 100s of MHz, and it has an SZ decrement at these frequencies, the radio SZ effect in combination with the CMB SZ signal would result in a null frequency $\nu_N$ between 700 and 800~MHz, below which there would be an increment in the observed surface brightness in the direction of a cluster and above which there would be a decrement, with the amount of increment or decrement larger at frequencies farther from the null.

In \citet{Lee22} we present further calculations of the relativistic and kinematic corrections to the combined RSB and CMB signals, as well as the potential effects of a dipole anisotropy in the RSB as seen by the cluster.  The magnitude of the combined radio SZ effect, and the exact frequency of the null, each depends to varying degrees on the electron temperature $T_e$ and density $N_s$ (often combined in the Compton $y$ parameter $y={{k_B T_e N_s}/ {m_e c^2}}$), the cluster velocity $\beta_c={{v_c}\over {c}}$ and movement direction $\mu_c=\cos{\theta_c}$, and the presence or absence of a dipole anisotropy in the RSB.  

As shown in \citet{Lee22} realistic models result in a combined signal null between $\sim$730< $\nu_N$<$\sim$800~MHz, and, assuming all of the RSB is present at the redshift of a given cluster, a decrease in the radiometric temperature of on the order of $\sim$0.25~mK at 1~GHz and an increase on the order of $\sim$1~mK at 500~MHz.  If one allows the fraction of the current RSB present at higher redshifts $f(z)$ to vary, these increment and decrement magnitudes depend nonlinearly on that fraction -- for example $f(z)=0.5$ results in a loss of around three-quarters of the radiometric temperature increase at 500~MHz. 

From CMB SZ measurements the locations of thousands of clusters are known, and some clusters have measured or constrained values for $y$, $\beta_c$, and $\mu_c$.  Several existing radio interferometric facilities in principle have the sensitivity to measure the level of increment or decrement in emission temperature due to the radio SZ effect.  This is discussed in \S \ref{conc}.  A detection of the radio SZ effect would confirm the RSB as extragalactic, and its presence or lack thereof in clusters of higher redshift would constrain $f(z)$ and therefore potentially the redshift(s) of origin of the RSB.

\subsection{Low-frequency Absolute Spectrum Distortions --- Jens Chluba}\label{Chluba}
Spectral distortions of the CMB have now been recognized as an important probe of the early-universe and particle physics \citep{Chluba2019BAAS,Chluba2021Exp}. It has been argued that the long-standing limits on the average energy release obtained with \COBEF \citep{Mather1994, Fixsen1996} could principally be improved by more than a factor of $\simeq 10^3$ using modern technology \citep{Kogut11, PRISMWPII}. This could provide a litmus test of \LambdaCDM and potentially even lead to the discovery of signals due to new physics \citep{Chluba2021Exp}.

When thinking about CMB spectral distortions, we frequently fall into the standard $\mu$ plus $y$-distortion picture \citep{Sunyaev1970mu, Sunyaev1970CoASP, Burigana1991, Hu1993}. However, not only has it become clear that the thermalization process allows for more rich signals when the distortion is created at the transition between efficient and inefficient Compton scattering around redshift $z\simeq 50,000$ \citep{Chluba2011therm,Khatri2012mix,Chluba2013Greens}.  It has also been demonstrated that photon injection distortions created at $z\lesssim 10^5$ generally can no longer be categorized using the simple $\mu/y$ picture
\citep{Chluba15,Bolliet21}. 

One prominent example of a photon injection distortion is the cosmological recombination radiation \citep{Dubrovich1975,Sunyaev2009,Chluba2016CosmoSpec}, which can tell us about the exact dynamics of the cosmological recombination process and potentially even allows measuring the cosmic expansion history at high redshift \citep{Hart2020,Hart2022}. At $z\lesssim 10^4$, Compton scattering is no longer efficient and photon injection processes essentially imprint the distortion signal that is only affected by redshifting and free-free absorption at low frequencies. The main motivation for thinking about the general photon injection problem was to try to understand if the distortion signals can mimic a power-law dependence at low frequencies due to the combined action of redshifting and scattering. Indeed, photon injection distortions in the low-frequency tail of the CMB, e.g., due to some decaying or annihilation particle, exhibit a rich phenomenology of spectral signals \citep{Chluba15,Bolliet21} that could be linked to the high RSB level inferred from ARCADE and other measurements discussed in \S \ref{intro}, if it is of cosmological origin. 

\begin{figure}
\centering
\includegraphics[width=\columnwidth]{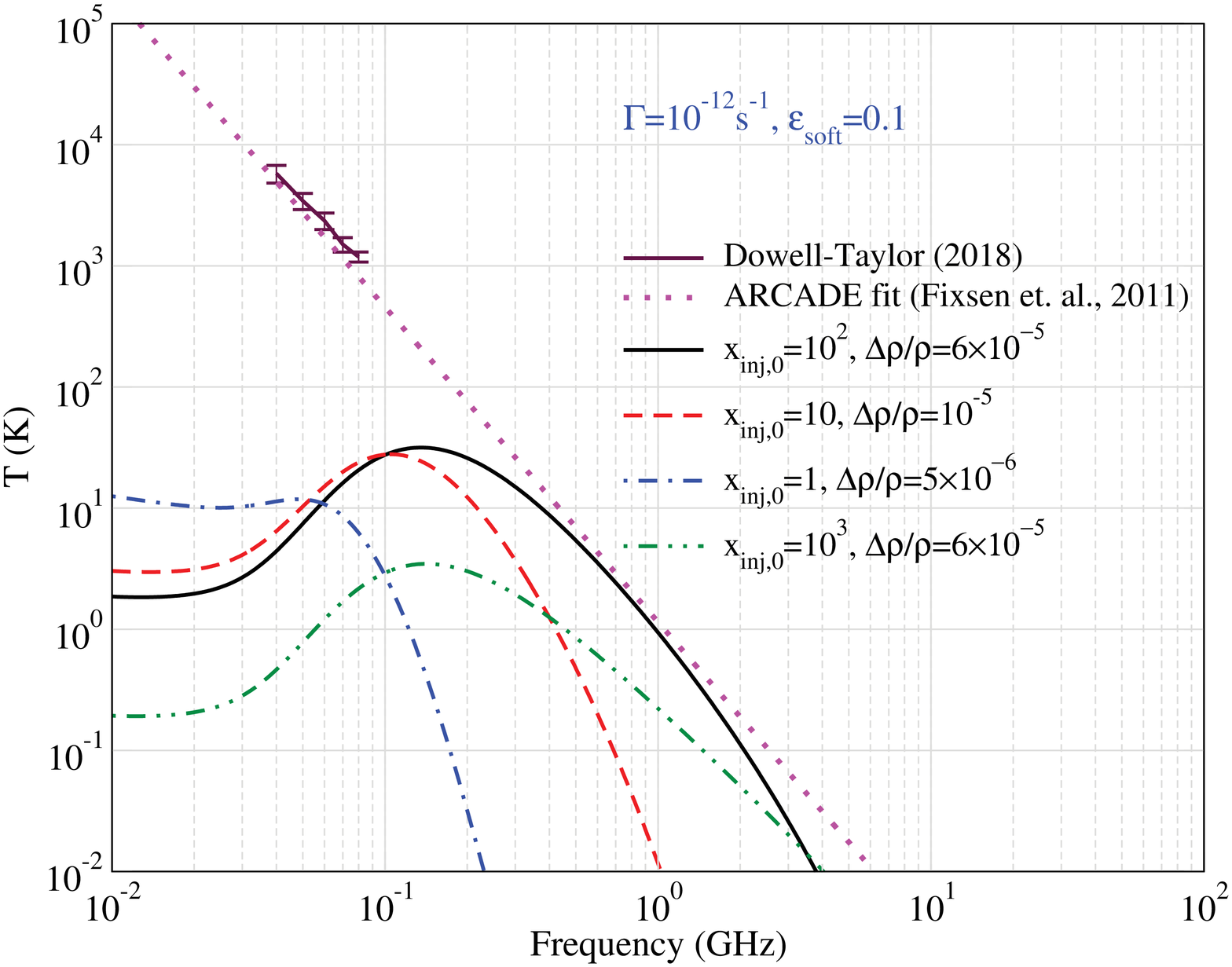}
\\[1mm]
\includegraphics[width=\columnwidth]{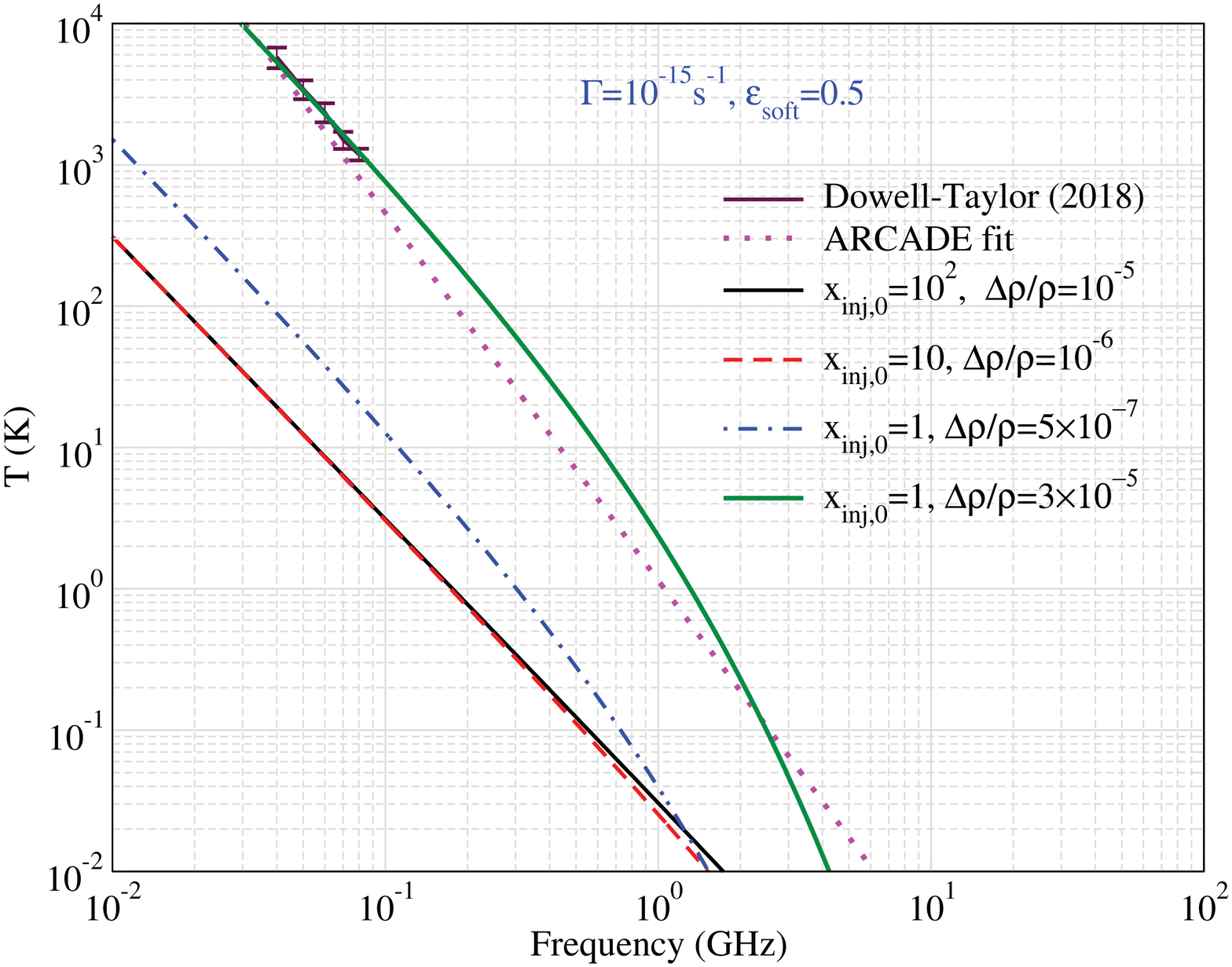}
\caption{Spectral distortion signals created by free-free-type soft photon injection from a decaying DM particle with lifetime $t_X\simeq 10^{12}$ (top) and $10^{15}$ s (bottom). In each case it is assumed that a fraction $\epsilon_{\rm soft}$ of the total energy (as labeled) is injected as soft photons. The injection frequency, $x_{\rm inj, 0}=h\nu_{\rm inj, 0}/k T_0$ is also varied with some of the cases shown coming intriguingly close to the ARCADE RSB level (figures provided by Sandeep Acharya).}
\label{fig:chluba}
\end{figure}
However, it appears that injection of photons at single frequencies may not be sufficient even if occurring in the partially Comptonized regime at $10^4\lesssim z \lesssim 3\times 10^5$ \citep{ACip}. We therefore considered more general photon injection cases with a power-law soft photon spectrum. In Fig.~\ref{fig:chluba}, we show a few distortion signals created by a decaying DM particle with varying lifetime and injection energy. This is to motivate that indeed it may be possible to create distortions at low frequencies by injecting soft photon spectra (here of free-free type) that come close to reproducing the high RSB level inferred from ARCADE and other measurements. Needless to say that these examples are just for illustration and a more rigorous search for viable solutions is currently in preparation.
Overall, it seems clear that new physics examples should consider the interplay with CMB thermalization and scattering physics to open the door to realistic predictions for the source of the RSB level.
An early-universe solution for the radio excess would also overcome limitations due to constraints on the fluctuations of the RSB (as discussed in \S \ref{Heston}), which other models (e.g. such as the one presented in \S \ref{Caputo}) may still suffer from \citep{Acharya2022}.

\subsection{Constraining Below-threshold Radio Source Counts with Machine Learning --- Elisa Todarello}\label{Todarello}

To determine whether there is a new population of faint point sources that give rise to the RSB, we try to develop a new technique to extract low flux density source counts from observational images based on Convolutional Neural Networks (CNN). Below-threshold source counts are usually determined through a statistical analysis of the confusion amplitude distribution, the so-called $P(D)$ method \citep{Scheuer}.

\begin{figure}
\centering
\includegraphics[width=0.4\textwidth]{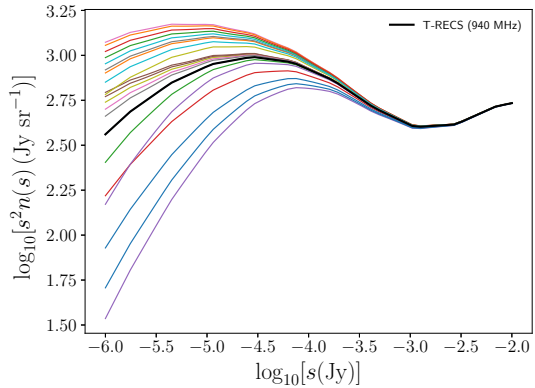}
\caption{Differential number counts used to create the training set discussed in \S \ref{Todarello}.}
\label{training_set}
\end{figure}

CNN's are well suited for image processing and have proven extremely powerful in pattern recognition. They are also used for counting tasks, such as determining the number of people in a densely packed crowd. It is then interesting to explore whether CNN's are able to outperform the $P(D)$ strategy, or at least to provide a complementary approach.

\begin{figure*}
\centering
\includegraphics[width=\textwidth]{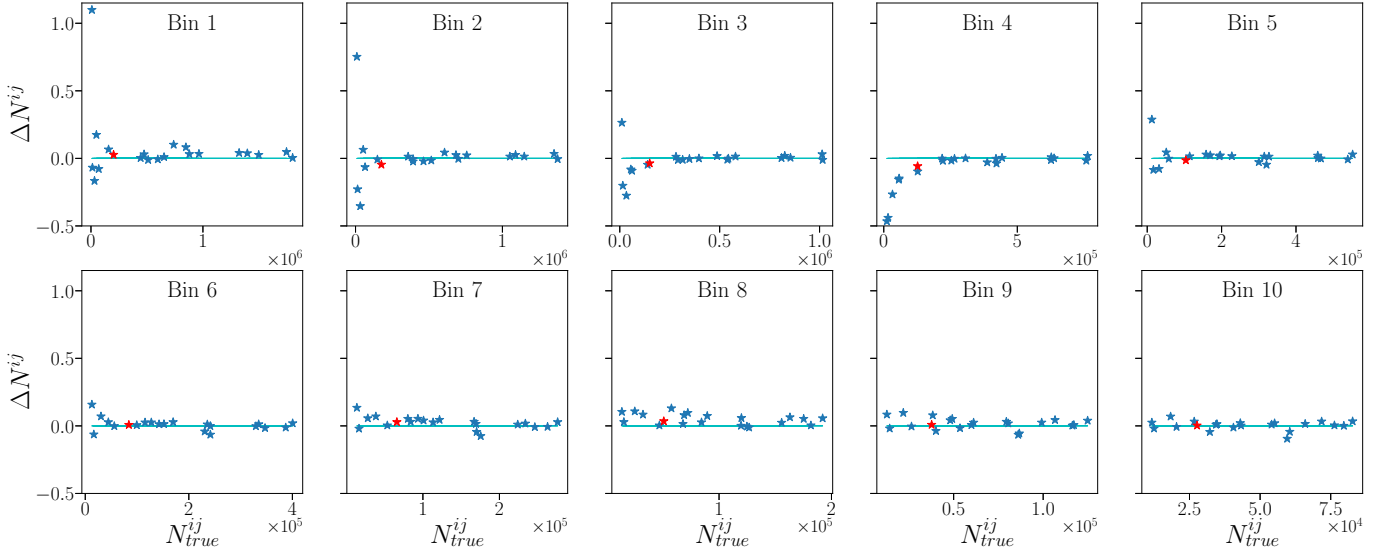}
\caption{Reconstruction residuals. The red star marks the image having the differential source count $n_{T-RECS}(s)$.}
\label{result}
\end{figure*}

Our goal is to train a CNN capable of inferring the source count at low flux densities $s$ from interferometric images, such as those of the Evolutionary Map of the Universe radio survey \citep[EMU -- ][]{Joseph19}.
Specifically, the output we want from the network is the source count in 10 logarithmically spaced flux bins between $10^{-5}$ and $10^{-7}$~Jy. Our first task is then to create a suitable training set of simulated images with known source counts. As a starting point, we take the Tiered Radio Extragalactic Continuum Simulation (T-RECS) simulated ``medium" catalog of extragalactic sources \citep{Bonaldi} at a frequency of 940~MHz. We truncate the catalog at a minimum flux of $10^{-7}$~Jy to render the file size manageable. This catalog spans 25~deg$^2$ and, with our truncation, contains about 30 million sources. The differential number count of sources $n(s)=dN(s)/ds$ reproduces observations. Next, we create new catalogs with a variety of $n(s)$ by modifying the T-RECS catalog. We choose the following functional form with two free parameters $\alpha$ and $s_0$
\begin{equation}
n(s) = \left( 1 + \frac{s_0}{s}\right)^\alpha n_{T-RECS}(s) \enspace.\label{ns}
\end{equation}
We consider 21 pairs of $\alpha$ and $s_0$ as shown in Figure~\ref{training_set}. 
We generate the 21 corresponding catalogs by Monte Carlo sampling $n(s)$. Several of these catalogs contain a number of sources greater than 30 million. We take the properties of the extra sources from the T-RECS ``wide" catalog, overwriting their coordinates with random values that fall within our image. 

To create the simulated images, we use ASKAPsoft \citep[Yandasoft -- ][]{Guzman}. In the first stage of the simulation, the text catalog is converted into a ``sky model", i.e. an image of the sky without telescope effects. Next, the observation is simulated, the output being the visibilities with instrumental noise added. In the last step, the visibilities are converted to physical space, and deconvolution with the point spread function is performed with the CLEAN algorithm.

At this point, we have 21 25~deg$^2$ images, each made of 2560$^2$ pixels. Since CNN's work more efficiently with low numbers of pixels, we split each image into 400 sub-images, for a total of 8400 sub-images, most of which we will use as a training set, while the rest will be used for validation and testing. Our CNN comprises three convolutional layers, and one densely connected layer before the output layer with 10 nodes, each corresponding to the source count on one of our bins.

As a first trial, we train the network using the sky model images. The network yields good results after about 500 epochs of training. Figure~\ref{result} shows
the reconstruction residual in the bin $i$ for the image $j$
\begin{equation}
\Delta N^{ij} =  \frac{N^{ij}_{predicted} - N^{ij}_{true}}{N^{ij}_{true}} \enspace,
\end{equation}
where the $N^{ij}$'s are obtained by summing over all sub-images in the training set that belong to the same 25 deg$^2$ image.

The worst performance is for images with few sources. However, such low source count are far from expectations.
We test that the network is able to reliably reconstruct the number counts for values of $\alpha$ and $s_0$ it hasn't seen before. As a stress test, we also create images for which the number count in each bin is assigned at random, within the range of values used for training. In this case, the network does not perform as well, indicating that it has learned the functional shape Eq.~\ref{ns} and it is not able to estimate the number of sources in each bin independently from the others. As a solution to this problem, we plan to retrain the CNN with a variety of physically plausible functional shapes, increasing the degrees of freedom from the current two, $\alpha$ and $s_0$.

The next and challenging step, that is currently in progress, is to apply the CNN to recover the source count from the restored image that contains noise and confusion.

\subsection{Observational Cosmology with the 21 cm Background Radiation (and Radio Background By-products) --- Gianni Bernardi}\label{Bernardi}

\begin{figure}
\includegraphics[width=1\linewidth]{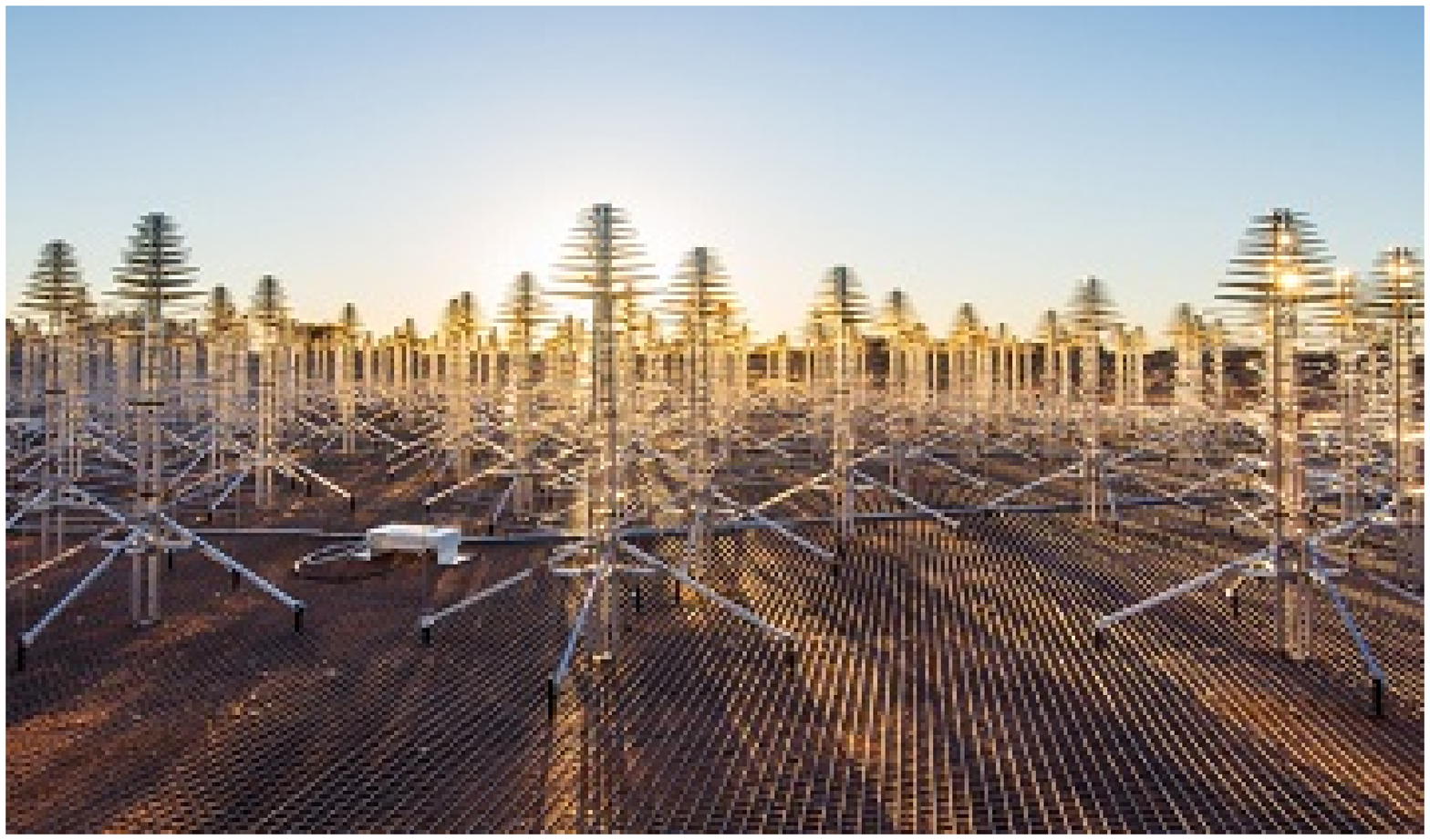} \\
\includegraphics[width=1\linewidth]{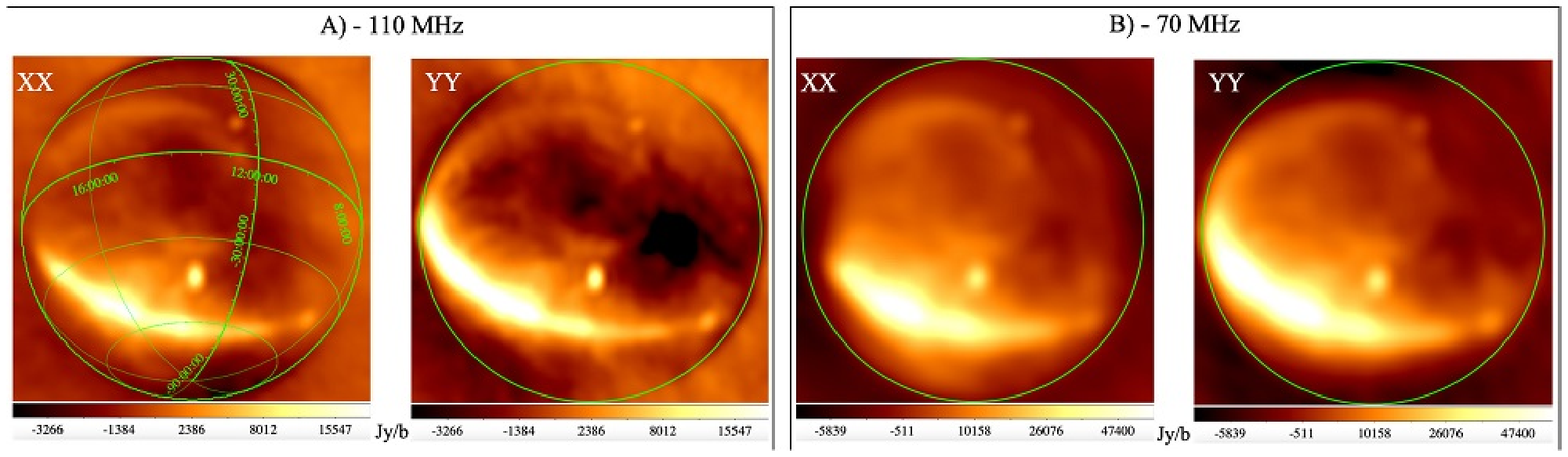} \\
\includegraphics[width=1\linewidth]{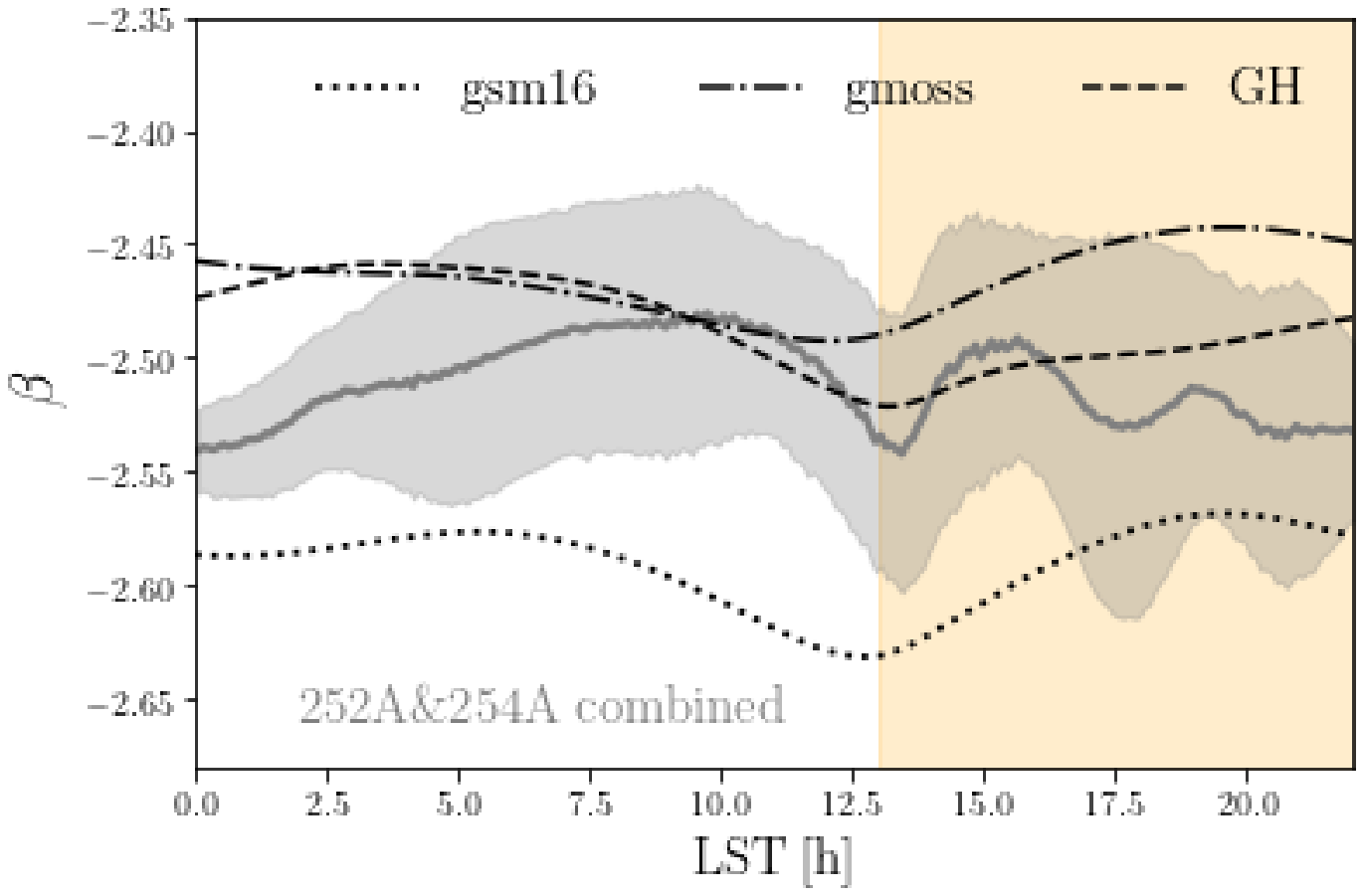}
\caption{{\bf Top}: Picture of the AAVS2 station with a close-up view of the  log-periodic antennas. Adapted from \cite{Macario22}.  {\bf Middle}: Examples of all-sky images taken with the AAVS2 station at 110 MHz (panel A) and 70 MHz (panel B). Both orthogonal polarizations, $XX$ and $YY$, are displayed. Units are Jy~beam$^{-1}$. Adapted from \cite{Macario22}.  {\bf Bottom}: Mean value (gray solid line) and standard deviation (gray area) of the foreground spectral index as a function of LST. Black lines represent prediction extrapolated from other measurements (see \citealt{Spinelli21} for details). The yellow region indicates measurement taken during daytime. Reproduced from \cite{Spinelli21}.}
\label{Bernardifig}
\end{figure}

The redshifted 21~cm line promises to be one of the best probes of the formation of early structures during the cosmic dawn and the subsequent epoch of reionization. This has motivated the construction of a new generation of radio instruments that are currently providing increasingly stringent upper limits on the expected signal \citep{Bernardi16,LOFAR,Trott20,HERA,Singh22,Barry22}, including a ``controversial" detection at $z \sim 17$ \citep{EDGES}. The challenge that 21~cm observations face are the separation of the cosmological signal from the much brighter foreground emission. The characterization of the foreground spatial and spectral properties has therefore been an active research line over the last decade \citep[e.g.][]{Bernardi10,Bernardi13,Dillon15,Thyagarajan15,Kerrigan18,Ghosh20,Garsden21,Cook22,Byrne22}. Such foreground characterization includes recent observations taken with two different instruments:
\begin{itemize}
    \item all-sky maps with the Aperture Array Verification System 2 \citep[AAVS2][]{Benthem21,Macario22} as pictured in the top panel of Figure~\ref{Bernardifig}. AAVS2 is a prototype station of the Square Kilometre Array, i.e. a $\sim 40$~m diameter station equipped with 256 dual-polarization, log-periodic antennas sensitive to sky emission in the $50 - 350$~MHz range. A set of snapshot observations, spanning the whole $0 - 24^{\rm h}$ local sidereal time (LST) range, was carried out in interferometric mode in April 2020 in order to commission the newly deployed system. Each snapshot yielded an all-sky image like the one showed in Figure~\ref{Bernardifig} (middle panel), with angular resolutions between $\sim 1.3^\circ$ and $\sim 8^\circ$. As the telescope was used in drift scan mode, images show the brightness distribution changing as the sky transits overhead.
    
    Despite the limited angular resolution due to the longest baseline being only ($\sim 40$~m), the $uv$ coverage is excellent, with baselines as short as $\sim 1$~observing wavelength, which corresponds to angular scales as large as hundreds of degrees on the sky. Figure~\ref{Bernardifig} (middle panel) shows an example of how the large scale emission is accurately imaged in AAVS2 observations: the Galactic plane is visible in its entirety and large-scale and fainter, low-surface brightness features are detected across the whole sky. We found that the calibration accuracy is within 20\%, and further analysis can improve it. Future work will be dedicated to include the zero-spacing in AAVS2 observations in order to use them to measure the radio spectrum at high Galactic latitudes similarly to \citet{DT18}.

    \item measurements of the Galactic synchrotron spectrum with LEDA \citep{Bernardi15,Bernardi16,Price18}. LEDA is located at the Owens Valley Radio Observatory and uses four dipoles equipped with custom-built receivers that enable accurate total power radiometry in the $30 - 88$~MHz range. Absolutely calibrated spectra are obtained every 15~seconds as a function of LST and two dipoles observed for 137 nights between 2018 and 2019 \citep{Spinelli21}. Each spectrum $T_m$ is fitted by a power law model:
    \begin{equation}
        T_m (\beta, T_{75}) = T_{75} \left( \frac{\nu}{75 \, {\rm MHz}} \right)^\beta + T_{\rm CMB},
    \end{equation}
    where $\beta$ is the synchrotron spectral index, $T_{75}$ is the sky temperature at 75~MHz, $\nu$ the observing frequency and $T_{\rm CMB}$ is the Cosmic Microwave Background temperature. We found that the spectral index is reasonably constant in the $0^{\rm h} < {\rm LST} < 22^{\rm h}$ range, varying between $-2.48$ and $-2.54$, with a tight dispersion $\Delta \beta \sim 0.06$  as seen in the bottom panel of Figure~\ref{Bernardifig} \citep{Spinelli21}). LEDA observations have an accurate absolute calibration and are sensitive to the whole sky emission visible from the Owens Valley Radio Observatory, including the Galactic plane. Future work will be dedicated to model the known (Galactic and extragalactic) contributions to the measured spectrum in order to constrain the contribution of the radio synchrotron background excess. 
\end{itemize}

\subsection{Backgrounds from Primordial Black Holes --- Nico Cappelluti}\label{Cappeluti}

In this section we explore the observational implications of a model in which primordial black holes (PBHs) with a broad birth mass function ranging in mass from a fraction of a solar mass to $\sim$10$^6$ M$_{\odot}$, consistent with current observational limits, constitute the DM component in the universe as presented by \cite{cap22}.  The formation and evolution of DM and baryonic matter in this PBH-\LambdaCDM universe are presented. 

In this picture, PBH DM mini-halos collapse earlier than in standard \LambdaCDM, baryons cool to form stars at $z\sim15-20$, and growing PBHs at these early epochs start to accrete through Bondi capture. The volume emissivity of these sources peaks at $z\sim20$ and rapidly fades at lower redshifts. As a consequence, PBH DM could also provide a channel to make early black hole seeds and naturally account for the origin of an underlying DM halo/host galaxy and central black hole connection that manifests as the $M_{\rm bh}-\sigma$ correlation. 

To estimate the luminosity function and contribution to integrated emission power spectrum from these high-redshift PBH DM halos, we develop a Halo Occupation Distribution (HOD) model. In addition to tracing the star formation and reionizaton history, it permits us to evaluate the Cosmic Infrared and X-ray backgrounds (CIB and CXB). We find that accretion onto PBHs/AGN successfully accounts for these detected backgrounds and their cross correlation, with the inclusion of an additional infrared stellar emission component. Detection of the deep infrared source count distribution by the James Webb Space Telescope (JWST) could reveal the existence of this population of high-redshift star-forming and accreting PBH DM.

Finally, by employing the formalism of \citet{has20}, we show that if a fraction of accreting PBHs similar to that observed in AGN in the local universe are radio loud, this model can easily reproduce the enhancement of radio background at high redshifts required to explain the EDGES 21-cm trough result, which is a fraction of the current RSB level in the universe. 

\subsection{Background of Radio Photons from Primordial Black Holes --- Shikhar Mittal}\label{Mittal}

\citet{FH18} first showed that a radio background can enhance the 21-cm signal and potentially explain the amplitude depth seen in the EDGES \citep{EDGES} measurement. We consider accreting primordial black holes (PBHs) as the originator of the RSB as discussed in \citet{Mittal_2022}.

PBHs are interesting DM candidates formed in the early universe by a gravitational collapse of overdense regions. They are predicted to exist over a wide range of masses. Current observations put constraints in the mass range $\sim10^{-18}$--$10^{21}\,\mathrm{M}_{\odot}$ \citep{carr}. Black holes of masses a few orders of magnitude higher than $\mathrm{M}_{\odot}$ are important for studying accretion phenomenon. These black holes are comparable in mass to the astrophysical supermassive black holes that reside in the centers of galaxies and power active galactic nuclei.

Accreting objects generate strong relativistic jets that span a wide range of frequencies in the electromagnetic spectrum. The synchrotron mechanism \citep{begelman} along with first-order Fermi acceleration \citep{Bell1, Bell2} predict the radio emissivity from accretion jets to follow a power law of index $\approx-0.6$. The resulting excess sky brightness temperature has a power-law dependence on frequency, $T_{\mathrm{b}}\propto\nu^{\beta}$, where $\beta=-2.6$, which is same as the index reported by ARCADE~2/LWA1. This makes radio-emitting accreting PBHs well-motivated candidates as the generator of the RSB.

\begin{figure}
\centering
\includegraphics[width=1\linewidth]{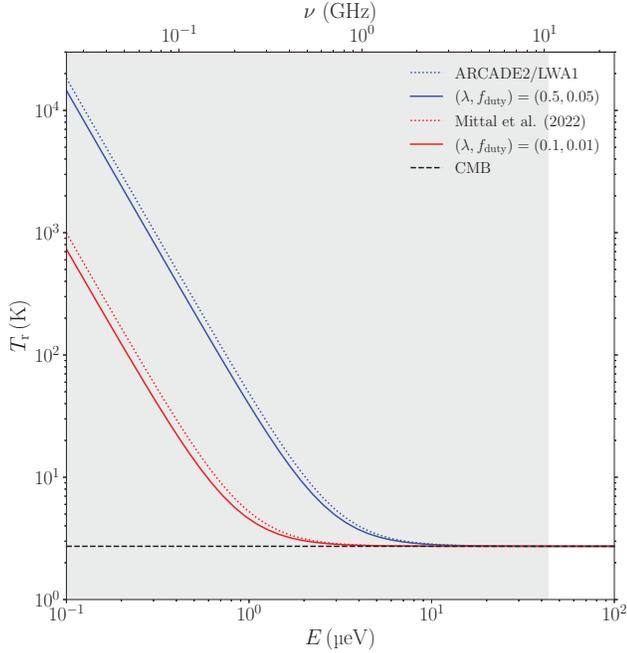}
\caption{The net background temperature $(T_\mathrm{r}=T_{\mathrm{b}}
+T_{\mathrm{CMB}})$ generated by radio emission due to accretion onto supermassive PBHs. For $\lambda=0.5, f_{\mathrm{X}}=0.1, f_{\mathrm{duty}}=0.05$ and $f_{\mathrm{PBH}}=10^{-4}$ (solid blue) we can explain the RSB observed (dotted blue). For $\lambda=0.1, f_{\mathrm{duty}}=0.01$ we get (solid red) 5\% of ARCADE~2 (dotted red) radio emission necessary to get the EDGES result \citep{Mittal_JCAP}. The CMB temperature is shown in dashed black for reference. The gray shaded region roughly covers the frequency range over which an RSB has been seen.}\label{Mittalfig}
\end{figure}

The number density of accreting black holes times the luminosity from a single accreting black hole gives an estimate of the total emissivity. Number density can be calculated, assuming a monochromatic mass function and a homogeneous distribution of PBHs, from the mass density of PBHs which in turn can be written as a fraction $f_{\mathrm{PBH}}$ of DM mass density. The single black hole radio luminosity can be calculated from an empirical relation, the so-called fundamental plane of black hole activity, that connects radio luminosity, X-ray luminosity and the black hole mass. One such relation is provided by \citet{Wang_2006} calibrated at a radio frequency of 1.4\,GHz and total X-ray luminosity for photon energies in the range $0.1$--2.4\,keV. In order to model the X-ray luminosity we assume that it is a fixed fraction ($f_{\mathrm{X}}\sim0.1$) of the bolometric luminosity which in turn is a fraction $\lambda$ (Eddington ratio) of the Eddington luminosity. Assuming that a probability $f_{\mathrm{duty}}$ (duty cycle) for the black hole to be actively accreting at a particular time, we have at least two free handles to change in order to get the correct radio brightness temperature. Our final expression for comoving radio emissivity due to accreting PBHs is
\begin{eqnarray}
\epsilon_{\mathrm{acc}}(E)=5.65\times10^{19}f_{\mathrm{duty}}(f_{\mathrm{X}}\lambda)^{0.86}\left(\frac{f_{\mathrm{PBH}}\rho_{\mathrm{DM}}}{1\,\mathrm{kg m}^{-3}}\right)
\\
\nonumber \times\left(\frac{E}{5.79\,\upmu \mathrm{eV}}\right)^{-0.6}\mathrm{s}^{-1}\mathrm{m}^{-3}\,,\label{rem}
\end{eqnarray}
where $\rho_{\mathrm{DM}}$ is the mass density of DM today. We sum the emission -- accounting for the cosmological redshift -- starting from the epoch photons have been propagating freely, which we assume to be the last scattering of the CMB, i.e., $z_0\sim1000$. The resulting radio background specific intensity is
\begin{equation}
J_{\mathrm{acc}}(E,z)=\frac{c}{4\pi}(1+z)^3\int_z^{z_0}\frac{\epsilon_{\mathrm{acc}}(E')}{1+z'}\,\frac{\mathrm{d} z'}{H(z')}\,,\label{jacc}
\end{equation}
where $E'=E(1+z')/(1+z)$ and $H$ is the Hubble function. As we are interested in observations made today we put $z=0$.

For currently the strongest constraint ($f_{\mathrm{PBH}}\sim10^{-4}$) obtained by dynamical effects (in particular by halo dynamical friction) on accreting supermassive PBH of mass $\sim10^{8}\,\mathrm{M}_{\odot}$ \citep{Carr_1999, carr}, and $\lambda=0.5, f_{\mathrm{duty}}=0.05$ \citep{Shankar_2008, Rai09} we get the net brightness temperature as shown by the blue solid curve in figure~\ref{Mittalfig}. The blue dotted curve shows the ARCADE~2 result (equation~\ref{T_B}). Within the uncertainties of the free parameter, for $\lambda=0.1, f_{\mathrm{duty}}=0.01$ we get 5\% of ARCADE~2 radio emission, which is necessary to obtain the depth in the EDGES measurement of 21-cm signal \citep{Mittal_JCAP}.\footnote{Along with an enhanced Lyman-$\alpha$ coupling \citep{Mittal_2020}, though not sufficient to explain the shape of EDGES profile as explained by \citet{Mittal_EDGES}.} The dotted red curve shows the level of radio background required for EDGES and the solid red curve shows the net brightness temperature from accreting PBHs for lower values of $\lambda$ and $f_{\mathrm{duty}}$. In both cases the solid and the dotted curves are in excellent agreement with each other as expected since the synchrotron radiation from jets follow a power law of index same as that reported by observations for frequencies in the radio band.

An obvious question for the scenario discussed here is whether it is allowed by constraints from measurements of the X-ray background and the constraints on reionization. Unfortunately, computing the contribution of the accreting PBHs discussed here to the X-ray background and reionization requires making several poorly understood assumptions all the way to $z\sim 1000$. Nonetheless, with a naive application of our low-redshift understanding of AGN specrtal energy distributions to the accreting PBHs, we find that the model can evade the X-ray constraints if the accreting PBHs have a radio-loud fraction similar to AGN. The accreting PBHs also evade reionization constraints if they have obscuration fractions similar to those of AGN.

\subsection{Can the Local Bubble Explain the Radio Background? --- Martin Krause}\label{Krause}

The Local Bubble is a low-density cavity in the interstellar medium around the Solar system \citep[e.g.,][]{CR87}, likely formed by winds and explosions of massive stars \citep{Breitschwea16,Schulrea18a}.  Hot gas in the bubble contributes significantly to the soft X-ray background \citep[e.g.,][]{Snowdea97,Snowdea98}.  The boundary is delineated by a dusty shell \citep{Lallea14,Pelgrea20} and
groups/associations of young stars \citep{Zuckea22}. 
The superbubble contains high ionization species \citep{BdA06}, filaments and clouds of partially neutral and possibly even molecular gas  \citep[e.g.,][]{GryJen17,RedLin08,RedLin15,Snowdea15,LRT19}
and is threaded by magnetic fields 
\citep[e.g.,][]{AP06,McComea11,Frischea15,Alvea18,Piirolea20}.
The leptonic cosmic ray distribution is directly measured with near-Earth detectors \citep[e.g.,][]{Aguilea19}.  The Local Bubble hence contributes to the radio synchrotron background.

As a guidance for the general distribution of the radio emission in the superbubble, one could take the nonthermal superbubble in IC10 \citep{Heesea15}, a smooth, round and filled structure without edge-brightening, that would produce the correct spectrum for the synchrotron background and 
more than enough flux when scaled to the Local Bubble.

Thanks to a number of measurements unique to the Local Bubble, it is possible to predict its radio emission fairly precisely.  Cosmic ray electrons are directly measured with the Alpha Magnetic Spectrometer (AMS) onboard the International Space Station (ISS) \citep{Aguilea19}.  Low energy cosmic rays are inhibited in their propagation through the Solar System by the magnetic field of the Solar wind. Constraints at lower energy and outside the volume influenced by the Solar wind by Voyager~I \citep{Cummea16} allow for the solution of the propagation problem and thus to derive the particle energy spectrum for the local interstellar medium, i.e., the Local Bubble \citep{Vittea19}: $n(E)\propto E^{-p}$, with $p=1.4$ (3.1) below (above) 1~GeV. 

The magnetic field in the local bubble is constrained by measurements of the Faraday effect \citep[e.g.,][]{XH19}.
Eight pulsars located near the edge of the Local Bubble,  all in one particular sector show a root mean square rotation measure  of 33~rad~m$^{-2}$ \citep{XH19}.  Their mean dispersion measure indicates  a column density of free thermal electrons of $N_e=(1.3\pm0.6)\times 10^{24}$~m$^{-2}$.
X-ray measurements of the hot bubble plasma suggest a thermal electron density of $n_{e,\mathrm{X}} = (4.68 \pm0.47) \times 10^3 $~m$^{-3}$ \citep{Snowdea14}, typical for superbubbles \citep{Krausea13a,Krausea14a}.  Warm clouds are observed within the Local Bubble. They have sizes of several parsecs and electron densities of the order of $n_{e,\mathrm{wc}}=10^5$~m$^{-3}$ \citep[e.g.,][]{GryJen17,LRT19}.  Pressure balance with the volume filling X-ray plasma generally suggest $\approx 0.5$~nT for warm clouds in the Local Bubble \citep{Snowdea14}.  Such data allows estimates of the rotation measure contributions \citep[][for details]{KH21}.  The bulk of the rotation measure
is clearly not contributed by warm clouds. It could come from unknown fractions from bubble wall and hot X-ray plasma inside the bubble, which limits the magnetic field to
about 10~nT (100~$\upmu$G).
\begin{figure*}[th]\centering
\includegraphics[width=0.9\textwidth]{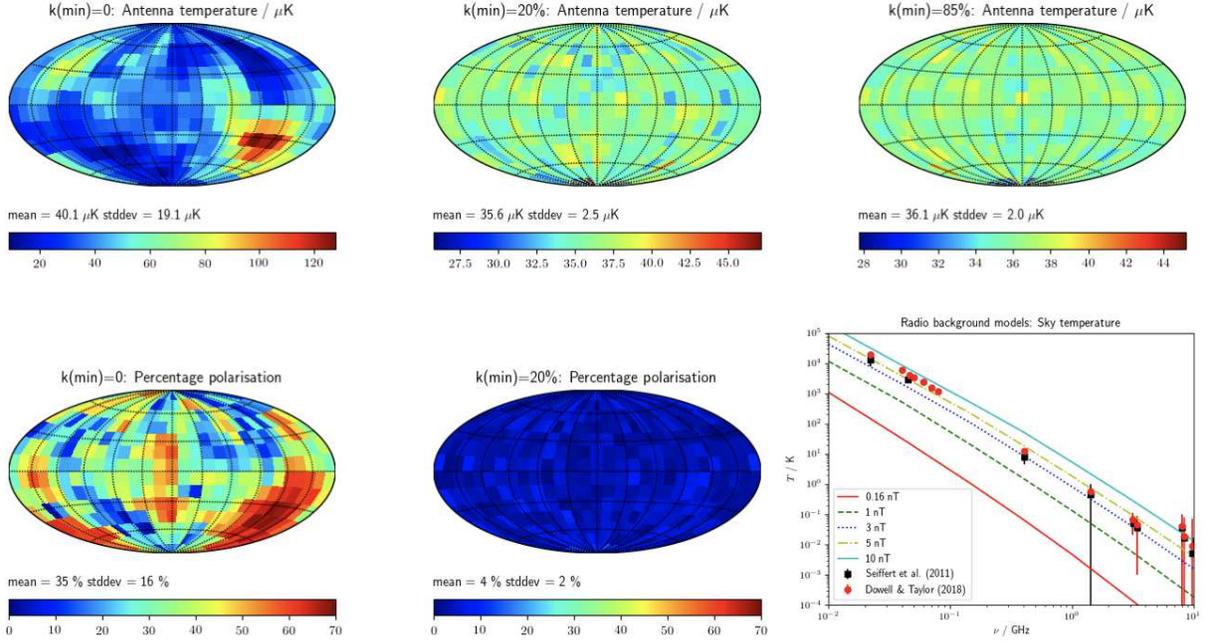}
\caption{Synthetic radio sky with a mean magnetic field of 1.6~nT at 3.3~GHz (except: bottom right). The resolution is 12$^\circ$ matching that of the ARCADE~2 radiometer.  The top row shows the distribution of the antenna temperature. The bottom row shows the fractional polarization for the corresponding image. The left column is for a complete Kolmogorov power spectrum. The middle (right) one is for a model with the 20 (85) percent largest modes set to zero.	Bottom right: Predicted radio synchrotron emission for the 	Local Bubble for the 85 percent largest modes set to zero and different mean magnetic field strengths between 0.16~nT   and 10~nT. Measurements are from \citet{Seiffert11} and \citet{DT18}. Reproduced from \citet{KH21}.}
\label{fig:skysim3GHz}
\end{figure*}

A normal turbulent cascade wipes out large-scale fluctuations of the magnetic field over time, unless the magnetic energy density dominates, in which case fluctuations on large scales are built up \citep[inverse cascade,e.g.,][]{CHB01,BKT15,Sur19}.  One can thus estimate the magnetic field strength from the geometry of the magnetic field, which is constrained by starlight polarization \citep{Berdea14,Pelgrea20}.  Some coherent large-scale structure seems to be associated with the edge, whereas the interior of the Local Bubble appears to have a magnetic field
structure characterized by decaying turbulence, with the largest magnetic filaments about 40 pc long.   If this is true, turbulence theory predicts that the magnetic energy density is not the dominant one in the Local Bubble (compare above).

The magnetic field strength for equipartition between magnetic and  thermal energy,  is $ B_\mathrm{eq,th}=0.61$~nT \citep{Snowdea14}.  Equipartition with the energy density in relativistic leptons in our Local Bubble model is reached for $B_\mathrm{eq,rel}=0.16$nT, a value that would also allow to interpret the break in the electron energy distribution at 1~TeV as due to synchrotron cooling \citep{Lopea18}.

Decaying turbulence is expected in the Local Bubble, as the last supernova happened about 1.5-3.2 Myr ago, as evidenced by deposits of radioactive $^{60}$Fe in deep sea sediments \citep{Wallnea16}. This corresponds to at least one sound crossing time through the Local Bubble, which is the characteristic decay time for turbulence. The picture is, however, complicated by the discovery of TeV electrons
which likely point to current energy injection by a pulsar wind \citep{Lopea18,Bykea19}.

We model the magnetic field in the Local Bubble as a random field with a vector potential drawn from a Rayleigh = distribution  with a Kolmogorov power
spectrum following, e.g., \citet{Tribble91} and \citet{Murgea04}.  We use magnetic field cubes with 256 cells on a side.  Following the experimental data on the field geometry, we set the 85 per cent largest modes to zero. 
We have also run models for the uncut power spectrum and for a cut at 20~\% for comparison. We put the observer in the center of the data cube, scale the magnetic field to values within the range allowed by observations and assume a homogeneous distribution of synchrotron-emitting leptons, distributed in energy space according to the model of \citet{Vittea19}. From this setup, we compute radio synchrotron flux and polarization according to standard formulae \citep[see][for details]{KH21}.

Synthetic antenna temperature and polarization maps are shown in Fig.~\ref{fig:skysim3GHz}.  There is little difference between the  sky distributions predicted for $k_\mathrm{min}=20$~\% and $k_\mathrm{min}=85$~\%. In both cases, the distribution is smooth across the sky with maximum antenna temperature ratios below two for any two sky directions and a standard deviation of less than 10 percent of the mean. A noteworthy polarization signal is only predicted for the full Kolmogorov power spectrum.

The Local Bubble has approximately a power law radio spectrum very similar to that of the radio background (spectral index $\alpha \approx 0.6$, Fig.~\ref{fig:skysim3GHz}, bottom right).  Good agreement with the RSB level is found for magnetic field strengths between 3 and about 5~nT.  While such a high magnetic  field would be allowed by the pulsar rotation measures,  as argued above, it would lead to a dominant magnetic energy density, hence to an inverse
cascade. Apart from the fact that such high magnetic fields are generally not observed in the interstellar medium, and that the expected large-scale fluctuations in the starlight polarization are not seen, this would also contradict
the observations of the low polarization of the RSB.

The most likely conclusion from this study is hence that the magnetic field in the Local Bubble is low, likely roughly in equipartition with the thermal energy density, or the one in relativistic particles. In this case, the Local Bubble can only contribute at the few percent level to the RSB.

\subsection{Synchrotron Polarization as a Test of the Radio Background --- Al Kogut}\label{Kogut}

The spectral dependence of the observed radio excess, $T_A \propto \nu^\beta$ with $\beta = -2.58 \pm 0.05$
\citep{Fixsen11,DT18} is nearly identical to known Galactic features and is highly suggestive of synchrotron emission.  Polarization provides a test of Galactic versus extragalactic origin. Synchrotron emission is inherently highly polarized; emission from a single isotropic region with a uniform magnetic field and cosmic ray energy distribution $N(E) \propto \kappa E^{-p}$ will have spectral index
$\beta = -(p+3)/2$ and fractional polarization $f = \frac{p+1}{p + 7/3}$ \citep{rybicki/lightman:1979}.
The observed spectral index implies  $p=2.2$ and fractional polarization as high as $f \sim 0.7$.

\begin{figure}[h]
\vspace{-8mm}
\includegraphics[width=3.2in]{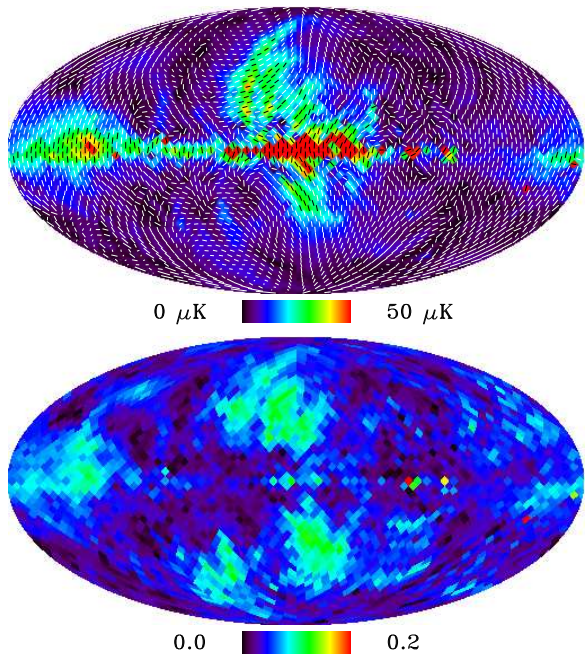}
\vspace{-8mm}
\caption{
Polarization intensity $P$ (top) and fractional polarization $P/I$ (bottom) from Planck evaluated at 30 GHz. Despite the high degree of depolarization, the polarization angle (white bars) shows little scatter.}
\label{KogutFig}
\end{figure}

The actual microwave sky, in contrast, is strikingly depolarized. Figure \ref{KogutFig} shows the polarized intensity $P = ( Q^2 + U^2 )^{1/2}$, polarization angle
$\psi = 0.5 \arctan( Q / U )$ and fractional [polarization
$f = P/I$ for the Planck synchrotron model 
\citep{planck_synch_I,planck_synch_QU} evaluated at a frequency of 30 GHz, where $I,Q,U$ are the Stokes parameters.  Two features are evident. Although synchrotron emission is nominally highly polarized, 50\% of the sky at Galactic latitude $|b|>20^\circ$ has fractional polarization $f < 0.031$. Despite this, the polarization direction is highly organized, with little scatter in polarization direction from neighboring pixels.  We quantify this by computing the difference in polarization angle
between each pixel and its neighbors, $\Delta \psi = \sum_j ( | \psi_i - \psi_j | )$ and find a median value
$\Delta \psi = 14^\circ$ at $|b| > 20^\circ$.

These two features are difficult to reconcile. Depolarization from multiple independent domains along individual lines of sight should also affect the polarization angle $\psi(\hat{n})$ in each line of sight $\hat{n}$. Since depolarization is a random process, the scatter in polarization angle from one line of sight to another should {\it increase} as additional independent domains align in different directions.

Magnetohydrodynamic (MHD) simulations allow a simultaneous assessment of both depolarization and alignment. We use the Enzo code \citep{Collins10,Bryan14} to generate realizations of a turbulent magnetic field within a $512^3$ data cube.  We begin with cubes of uniform density  and constant magnetic field $\vec{B}_0=B_0 \, \hat{x}$, then add kinetic energy on large scales. The MHD equations cascade the energy from the large injection scales to progressively smaller-scale structures, tangling the magnetic field in a physically accurate fashion \citep{Collins10}.
The ratio of the mean magnetic field to tangled magnetic field is determined by the ratio of thermal, kinetic, and magnetic energies, which we vary over a grid in the sonic Mach number $M =v_{\rm rms}/c_s$ and the Alfv\'{e}n Mach number $M_A =v_{\rm rms}/v_A$, with $c_s$ the speed of sound in the medium and $v_A$ the speed of an Alfv\'{e}n wave in the medium.  For each pair of Mach numbers we use the resulting 3-D magnetic field to evaluate the synchrotron emission vector within each cube cell, then sum the emission along each of the cube faces to evaluate the projected synchrotron emission  in polarization and intensity perpendicular and parallel to the input field.

\begin{table}[t]
\caption{MHD Simulation Results} 
\label{KogutTable}
{
\small
\begin{center}
\vspace{-2mm}
\begin{tabular}{|c c | c c | c c| }
\hline 
\multicolumn{2}{| c}{Mach Number} 		& 
\multicolumn{2}{| c}{Parallel to $\vec{B}_0$}	&
\multicolumn{2}{| c |}{Perp. to $\vec{B}_0$}	\\
$M$	& $M_A$ & $f$ & $\Delta \psi$ ($^{\circ}$) & $f$ & $\Delta \psi$ ($^{\circ}$)\\
\hline
0.5  &  0.5    & 0.33  &  40.0 &  0.68 &   1.6      \\
0.5  &  2.0    & 0.06  &  49.0 &  0.09 &  11.0      \\
1.0  &  0.5    & 0.34  &  43.0 &  0.69 &   1.7      \\
1.0  &  2.0    & 0.10  &  40.0 &  0.13 &  10.0      \\
2.0  &  2.0    & 0.17  &  38.0 &  0.23 &   9.0      \\
3.0  &  2.0    & 0.17  &  42.0 &  0.21 &   9.5      \\
\hline
\multicolumn{6}{| c |}{Planck~(nominal) ~~~~ $f$=0.031 and $\Delta \psi=14.1^{\circ}$} \\
\hline
\multicolumn{6}{| c |}{Planck~(corrected$^a$) ~~ $f$=0.144 and $\Delta \psi=14.1^{\circ}$} \\
\hline
\multicolumn{6}{l}{$^a$After removing monopole component (see text)} \\
\end{tabular}
\end{center}
}
\vspace{-8mm}
\end{table}

Table \ref{KogutTable} shows the results.   None of the simulations simultaneously reproduce both the Planck fractional polarization and angular scatter.  The lowest fractional polarizations from the simulations ($f = 0.06$) are seen for the face parallel to the ordered field, but the corresponding angular scatter for this face
($\Delta \psi = 49^\circ$) shows a near-complete lack of angular correlation.  Conversely, simulations viewed along the perpendicular faces show angular scatter roughly compatible with the Planck model, but now with fractional polarization  a factor of 3 or more higher than observed.

An extragalactic radio background can mitigate this tension.
If some or all of the observed radio monopole component is extragalactic, it should be removed from the Galactic synchrotron model prior to computing the fractional polarization.  The Planck synchrotron model applies a correction for the integrated emission of extragalactic radio sources, but this accounts for less than a quarter of the total radio monopole component.  Subtracting the entire radio monopole component from  the Planck model of unpolarized synchrotron emission increases the median fractonal polarization without affecting the scatter in polarization angles. The median values 
$f = 0.14$ and $\Delta \psi = 14\deg$ after monopole subtraction are now compatible with the range of MHD simulations.

\subsection{Strategies to Identify the Galactic Foreground --- Isabella P. Carucci}\label{Carucci}

This talk summarized the strategies for identifying an astrophysical background in \ion{H}{1} Intensity Mapping (IM). The contribution to the workshop discussion was twofold. i) To present the \ion{H}{1} IM observable, which constitutes a novel kind of data in the radio band. ii) To illustrate the ongoing efforts to characterize the foregrounds of \ion{H}{1} IM, focusing on the statistical techniques that are crucial in the characterization of weak signals and with promising applications to RSB science.

\ion{H}{1} resides abundantly in all galaxies and shines in the 21-cm line hyperfine transition. Therefore, it is a perfect candidate for mapping the universe's large-scale structure. However, the line's weakness prohibits the use of traditional surveys targeting galaxies for covering large areas and deep redshifts necessary for cosmological studies. To overcome this issue, we use the IM technique: instead of resolving individual sources, we collect all their integrated emission, scanning large portions of the sky quickly and economically and, at the same time, preserving the accurate redshift information from the 21-cm spectral line.

\ion{H}{1} IM is an emerging science area; many are the new or planned instruments that can perform such surveys, such as the Canadian Hydrogen Intensity Mapping Experiment \citep[CHIME -- e.g.][]{Mandana22},  Baryon Acoustic Oscillations from Integrated Neutral Gas Observations \citep[BINGO -- e.g.][]{Abdalla22}, Hydrogen Intensity and Real-time Analysis eXperiment \citep[HIRAX -- e.g.][]{Crichton22}, Tianlai \citep[e.g.][]{Fengquan21}, MeerKAT \citep[e.g.][]{Wang2021}, and eventually SKAO. \ion{H}{1} IM will bring new radio data in the 100~MHz -- 1~GHz regime. Although no experiments plan to achieve absolute calibration as \ion{H}{1} IM looks after temperature fluctuations, new radio data on a significant fraction of the sky means further information on, e.g., the diffuse galactic synchrotron radiation \citep{Irfan22}, the galactic magnetic field, and extragalactic emissions.

Despite its promising science, \ion{H}{1} IM measurements face formidable challenges, as shown in its first application in cross correlation with galaxies \citep{Chang10} -- no direct detection has been performed yet. The main reason is the presence of not-well-characterized contaminants: foregrounds of astrophysical origin are orders of magnitude more intense than the signal \citep{Alonso14}. 
Moreover, this substantial difference in intensity among components translates any possible tiny leakage due to the instruments' imperfections and calibration uncertainties into catastrophic contamination, which is hard to model or prevent \citep[e.g.][]{Shaw15}. 

Given the gravity of the cleaning problem, the community addresses it with Blind Source Separation (BSS) methods. The two main BSS strategies are Principal Component Analysis (PCA) and Independent Component Analysis (ICA) -- i.e., assuming the signal components are uncorrelated or statistically independent. Their application to data has been successful, especially on cross-detection with a second galaxy catalog \citep[e.g.][]{Masui13,Switzer13,Wolz22,Cunnington22}. However, we are improving the standard BSS techniques and optimizing them to our specific scope. 
In particular, we have recently devised a new component separation method named mixGMCA. It is a hybrid PCA and sparsity-based Generalized Morphological Component Analysis \citep[GMCA -- e.g.][]{Bobin07} BSS algorithm that uses wavelet decomposition of the maps and different treatments of their large and small spatial scales. GMCA assumes that the foreground components verify two hypotheses: sparse in a given domain (i.e., most samples are zero-valued) and morphologically diverse (their non-zero samples appear at different locations), easing their separation, which we achieve through a minimization problem. \citet{Carucci20} demonstrated the wavelets to be optimal to describe the \ion{H}{1} IM foreground components sparsely and highlighted how the sparse assumption holds better at small rather than large spatial scales. Since the wavelet decomposition offers a direct framework for analyzing multiscale data, we started allowing different mixtures and components to describe the signal at different scales. In practice, mixGMCA applies PCA on the largest scale\footnote{The coarse approximation of the maps resulting from the initial low-pass filtering of the wavelet decomposition.} and GMCA on the remaining ones. Then, it assembles the two solutions before re-transforming the maps into pixel space. mixGMCA participated in the first Blind Foreground Subtraction Challenge of the SKA Cosmology Science Working Group \citep{Spinelli22}. Results are promising: mixGMCA effectively removes the brightest diffuse astrophysical emissions with PCA while carefully handling the small-scale instrument-driven defects in the maps with GMCA. We are now testing mixGMCA on MeerKLASS data \citep{Wang2021}.

Component separation techniques are ubiquitous in data analysis. In particular, sparsity-based statistical learning algorithms have already been successfully applied in several astrophysical contexts, for example, CMB \citep{Bobin14}, high-redshift 21-cm interferometry \citep{Patil17}, and X-ray images of supernova remnants \citep{Picquenot19}. The component separation efforts we are conducting in the \ion{H}{1} IM context could be significant in the  measurements proposed during the workshop: i) the radio SZ effect (discussed in \S \ref{Holder}), and ii) cross-correlation analyses (discussed in \S \ref{conc}). Indeed, regarding the SZ effect, characterizing the halos, radio relics, and all structured components from the background is challenging, with higher chances of being achieved through the appropriate statistical methods as those here overviewed.

\subsection{Characterization of the Diffuse Radio Sky with EDGES and MIST --- Raul Monsalve}\label{Monsalve}

Understanding the origin of the radio synchrotron background requires accurate absolutely calibrated sky maps made at many frequencies over a wide frequency range. This talk focused on the efforts from the EDGES and MIST experiments to provide accurate absolute calibration to existing maps below $160$~MHz, which complements the work to produce new maps with intrinsically better calibration (Sections~\ref{Singal} and \ref{Bordenave}). The talk also described other ways in which EDGES and MIST are contributing to the characterization of the radio sky. 

\noindent \emph{EDGES}: The Experiment to Detect the Global EoR Signature (EDGES)\footnote{\href{https://loco.lab.asu.edu/edges/}{\url{https://loco.lab.asu.edu/edges/}}} employs  single-antenna, wideband, total-power radiometers measuring over the range $45-200$~MHz to attempt to detect the highly redshifted 21-cm signal from the cosmic dawn and the Epoch of Reionization at redshifts $30 \gtrsim z \gtrsim 6$. EDGES is located at the Murchison Radioastronomy Observatory (MRO) in Western Australia at a latitude of $-26.7^{\circ}$. Since 2008 EDGES has contributed to the characterization of the diffuse Galactic and extragalactic synchrotron radiation, which is the main foreground in measurements of the high-redshift 21-cm signal \citep{rogers2008,mozdzen2017,mozdzen2019}.

In \citet{monsalve2021} we reported the absolute calibration of two low-frequency maps using EDGES measurements. The maps are: 1) from \citet{guzman2011} at $45$~MHz, produced by combining the original maps by \citet{Alvarez97} at $45$~MHz and \citet{Maeda99} at $46.5$~MHz; and 2) from \citet{landecker1970} at $150$~MHz, which combines their own $150$-MHz observations with maps from \citet{yates1967} at $85$~MHz and \citet{turtle1962} at $178$~MHz. To calibrate the $45$-MHz map we used data from two EDGES low-band instruments \citep{EDGES} which, although are identical in principle, are observed on the sky with their dipole antennas at azimuthal orientations $\approx90^{\circ}$ apart. The absolute noise temperature calibration of EDGES is provided by laboratory measurements of resistive loads at $\approx300$~K and $400$~K connected at the input of the instrument in place of the antenna. To obtain our results we simultaneously fitted both EDGES low-band data sets to simulated observations produced by convolving the \citet{guzman2011} map with models of the EDGES low-band antenna beam. Our model for the calibrated map is $k_1\times {\rm input map} + k_2$, and the best-fit parameters found for the the \citet{guzman2011} map are $k_1=1.076\pm0.034$~$(2\sigma)$ and $k_2=-160\pm78$~K~$(2\sigma)$. Our results account for systematic uncertainties from receiver calibration, antenna orientation, ground properties, ionospheric and tropospheric effects, as well as from the choice of using a single data set in the analysis instead of both simultaneously. When this calibration is applied to the map, the brightness temperature at the reference Galactic coordinates $(l, b) = (+190^{\circ}, +50^{\circ})$ ---  in the region of lowest temperature in the sky --- goes from $3326 \pm \gtrsim 333$~K in the input map to $3419 \pm 255$~K~$(2\sigma)$ in the calibrated map. To calibrate the \citet{landecker1970} $150$-MHz map we used data from EDGES high-band \citep{mozdzen2017} and mid-band, whose horizontal blade dipole antennas are orientated $\approx90^{\circ}$ apart. Beyond the antenna orientations, the high- and mid-band systems are different in antenna size, ground plane size and shape, bandwidth, and receiver electronics. The results for this map are $k_1=1.112\pm 0.023$~$(2\sigma)$ and $k_2=0.7\pm6.0$~K~$(2\sigma)$, and the brightness temperature of the map at Galactic coordinates $(l, b) = (+190^{\circ}, +50^{\circ})$ goes from $148.9\pm \gtrsim 41$~K in the input map to $166.3 \pm 14.3$~K~$(2\sigma)$ in the calibrated map. 

Currently, we are working to provide absolute calibration to the $159$-MHz map recently published by \citet{kriele2022} from data obtained with the EDA2 SKA-Low prototype. For this, we are using observations with the same EDGES high- and mid-band blade dipole antennas as for the \citet{landecker1970} map, and also observations from an earlier version of the high-band instrument which measured with a Fourpoint antenna \citep{mozdzen2016}. Preliminarily, we find that the results from each of the individual data sets are very consistent, which represents strong validation of our understanding of the different instruments (R. Monsalve et al., in prep). EDGES has also recently reported the detection of H$\alpha$ and C$\alpha$ radio recombination lines (RRLs) in sky-average observations with the low-band and mid-band instruments. The detections occur across the full range of sidereal time although, as expected, their absolute amplitudes decrease toward high Galactic latitudes. Of particular interest is the report of the upper limit $33\pm11$~mK~$(1\sigma)$ for the absorption amplitude of the C$\alpha$ line at high Galactic latitudes when stacking all the expected RRL frequencies in the range $50-87$~MHz.

\begin{figure}
\includegraphics[width=3.3in]{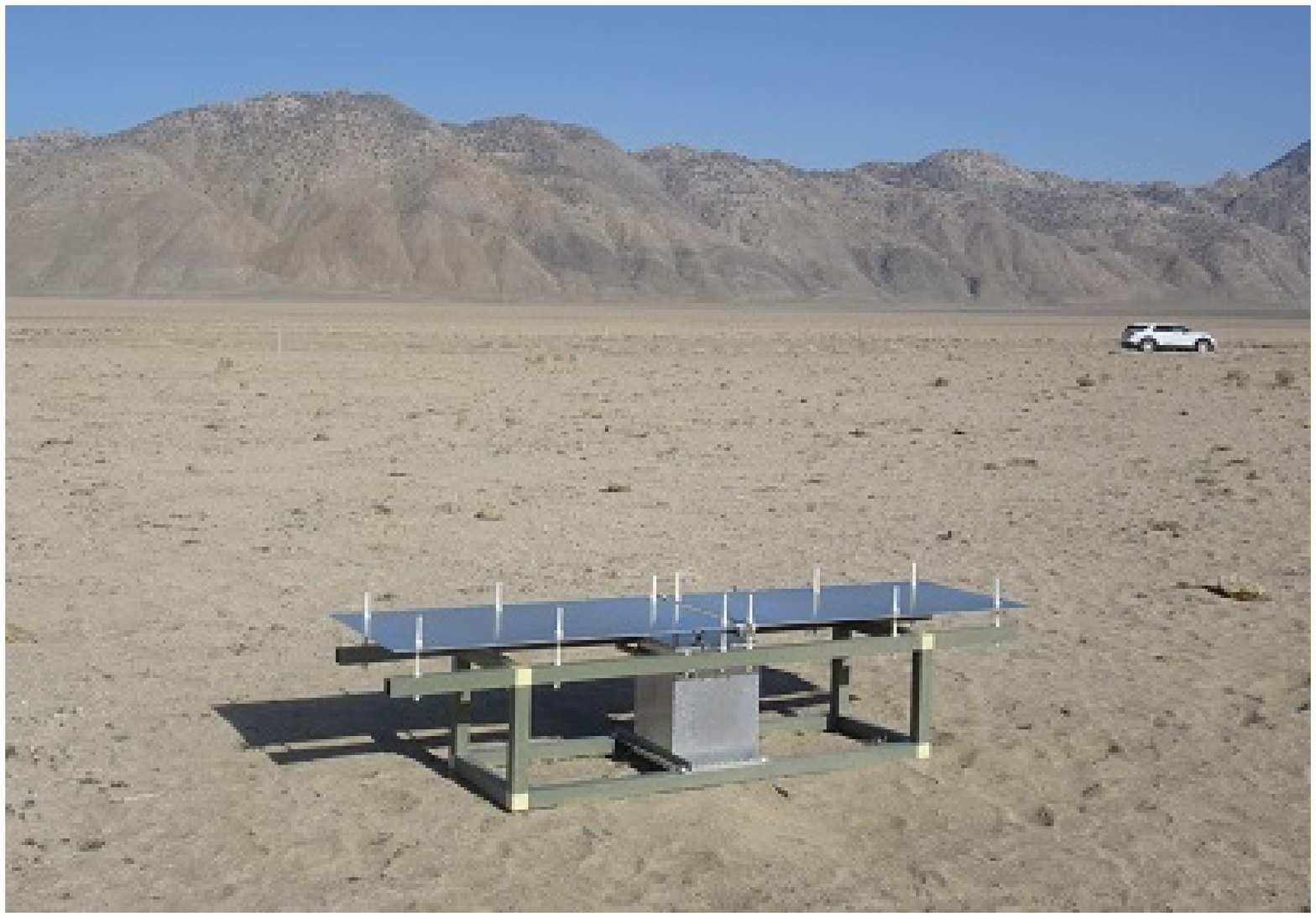}
\includegraphics[width=3.3in]{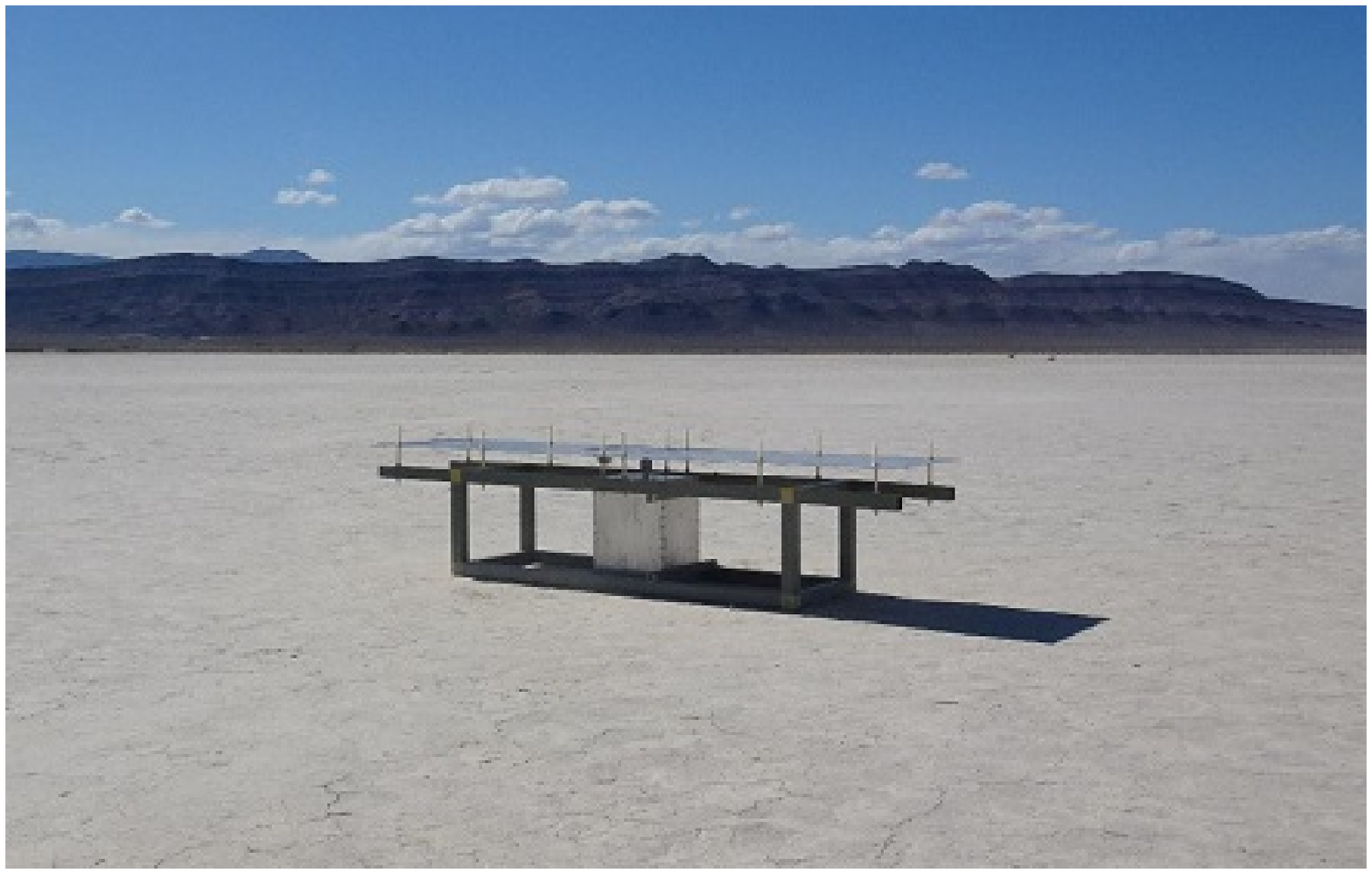}
\caption{MIST instrument conducting test observations in Deep Springs, California (top), and in Death Valley, Nevada (bottom) in May 2022. MIST is a single-antenna total-power radiometer without a metal ground plane. The antenna is a blade dipole of $2.4$ m tip-to-tip length and the measurements span the $25-125$~MHz range. These test measurements are currently being analyzed to understand the diffuse synchrotron emission in the northern hemisphere as well as the effect of the ground on the sky measurements.}
\label{MonsalveFig}
\end{figure}

\noindent \emph{MIST}: The Mapper of the IGM Spin Temperature (MIST)\footnote{\href{http://www.physics.mcgill.ca/mist/}{\url{http://www.physics.mcgill.ca/mist/}}} is a new experiment to detect the global high-redshift 21-cm signal, which has been designed to be portable and power efficient in order to conduct observations from several remote locations around the world. The instrument corresponds to a single-antenna total-power radiometer which, although similar, is different from EDGES in several ways. The main difference is that MIST does not use a metal ground plane in order to avoid potential systematic effects due to unaccounted-for interactions with the antenna and the soil. In May 2022, test measurements with MIST were conducted from the Deep Springs Valley, California, and the Death Valley, Nevada (see Figure \ref{MonsalveFig}). These measurements were very successful and are currently being analyzed to determine the contribution of the sky-averaged radio synchrotron emission in the northern hemisphere as well as the full effect of the ground on the sky measurements. Finally, given the success of the California and Nevada measurements, MIST planned to conduct observations from the Canadian High-Arctic, at a latitude of $\approx +79.5^{\circ}$, in July 2022. This site is excellent in terms of its radio-quiet conditions \citep{dyson2021} as well as in terms of the characteristics of the ground.

\subsection{Impact of Radio Background on the High-redshift 21-cm Signal --- Anastasia Fialkov}\label{Fialkov}

The redshifted 21-cm signal of neutral hydrogen \citep[e.g.][]{Furlanetto_2006, Barkana_2016, Mesinger_2019} can be described as a spectral distortion of the radio background spectrum which is typically assumed to be the cosmic microwave background (CMB). This signal is imprinted onto the radio background as it propagates through neutral high-redshift hydrogen gas. Radio photons at the intrinsic 1.42 GHz frequency are absorbed/emitted by hydrogen atoms if the characteristic temperature of the 21-cm transition, $T_S$, is colder/hotter than the temperature of the background radiation, $T_{\rm rad}$. If the radio background has an excess contribution in addition to the CMB, this will be manifested in the shape of the spectral distortion. Therefore, the 21-cm signal can be used to test cosmic origin of the observed RSB \citep[e.g. ][]{Fialkov_19, Reis_20}. The distortion is imprinted at the intrinsic frequency, corresponding to the wavelength of 21 cm, but  due to the expansion of the universe it is observed at a frequency lower by a factor $1+z$ where  $z$ is the redshift of the hydrogen cloud. For cosmological redshifts corresponding to the Epoch of Reionization (EoR) and cosmic dawn, $z\sim 5-50$, the signal is imprinted at frequencies below 200 MHz and  can be detected by low-frequency radio telescopes such as HERA \citep{HERA}, LOFAR \citep{LOFAR}, MWA \citep{MWA}, EDGES \citep{EDGES}, SARAS \citep{Singh_2019,Singh22}. 

First proposed by \citet{FH18}, the impact of the the excess radio background on the 21-cm signal was extensively modeled in later works \citep{EW18, Fialkov_19, MF19, Reis_20}. The main effects include  a  boost in the contrast between $T_S$ and $T_{\rm rad}$ which leads to a deeper 21-cm absorption trough, a weaker  coupling between $T_S$ and the gas kinetic temperature, $T_K$,  and an enhancement of the radiative heating \citep[e.g.,][]{Fialkov_19}.   Effectively, excess radio background over the CMB slows down the cosmic clock of the 21-cm signal. With an excess radio background it takes longer for the astrophysical processes (described below) to affect the 21-cm signal \citep{Reis_20}. Therefore, an enhanced radio background might have important implications for 21-cm cosmology.  

The signature of the excess radio background is partially degenerate with astrophysical effects. The 21-cm signal is a probe of the thermal and ionization states of the intergalactic medium (IGM) and, thus,  can be used to constrain the formation of the first stars and galaxies  \citep[see e.g.][]{Furlanetto_2006, Barkana_2016, Mesinger_2019}. As these first sources of radiation appear, their light affects the 21-cm signal via a number of processes. Arguably the most important effect is coupling of $T_S$ to   $T_K$  by stellar Ly-$\alpha$ photons \citep[the Wouthuysen-Field, WF,  effect,][]{Wouthuysen_1952, Field_1958}. It is owing to this effect that  the 21-cm signal is observable against the background radiation. The strength and exact redshift of the signal imprinted by the Ly-$\alpha$ photons depends on the properties and formation time of the very first stars.

The next major component that determines the shape of the distortion  is the gas kinetic temperature. At cosmic dawn  the thermal evolution is driven by adiabatic cooling, $T_K$ is lower than $T_{\rm rad}$ and the 21-cm signal is seen in absorption. X-ray radiation produced by the first population of X-ray binaries heats up the gas. We assume that X-ray luminosity scales with star formation rate and introduce a free scaling parameter $f_X$ that can be constrained with data. If as a result of this heating  $T_K$ exceeds  $T_{\rm rad}$, the 21-cm signal will be seen in emission at the EoR \citep{Fialkov_2014}.  Ly-$\alpha$ and radio photons also contribute to heating the gas \citep{Venumadhav_2018, Reis_2021}, but in most plausible scenarios their contribution is subdominant to that of X-rays.
 
In our recent works we have considered two types of excess radio backgrounds and their impacts on the 21-cm signal. In \citet{Fialkov_19} we focused on a phenomenological synchrotron radio background (motivated by the ARCADE~2 and LWA1 measurements) which adds a homogeneous  contribution to the high-redshift  $T_{\rm rad}$ that decays with time, while in \citet{Reis_20} we considered a radio background produced by a growing population of radio galaxies. Such a radio background exhibits fluctuations that trace the nonhomogeneous star formation. The strength of radio emission produced by a star forming region scales with star formation rate multiplied  by a free scaling factor $f_{\rm Radio}$ which can be constrained using data. The average value of $f_{\rm Radio}$ in star forming galaxies today is $f_{\rm Radio} = 1$. We calculate the temperature of the radio background by integrating over the contribution of all galaxies within the past light-cone. 

For extreme values of $f_{\rm Radio}$ \citep[$f_{\rm Radio}> 300$ e.g.][]{Reis_20}, the excess radio background created by galaxies can explain the anomalously deep (and yet unverified) sky-averaged 21-cm signal reported by the EDEGS collaboration \citep{EDGES}. The strong radio background needs to be accompanied by strong X-ray heating and Ly-$\alpha$ background to ensure a narrow absorption profile similar to the reported one.   Even though the EDGES detection has not been confirmed \citep[e.g.][]{Hills_2018,Singh22}, the presence of the strong radio background at high redshifts  can be tested with other 21-cm experiments that provide upper limits including  the interferometers HERA \citep{HERA}, LOFAR \citep{LOFAR} and MWA \citep{MWA} as well as global signal experiments such as SARAS2 \citep{Bevins_2022} and SARAS3 \citep{Bevinsprep}. 

Out of the existing limits on the power spectrum, the strongest constraint comes from the HERA measurement at $z=8$.  Comparing our models to this limit we find constraints on the strength of the radio background and X-ray heating  \citep[see Section 8 of][]{HERA}. We rule out (with 95\% confidence) the combination of high radio luminosity of high-redshift galaxies of $L_{r,150{\rm MHz}}/SFR > 4 \times  10^{24}$ W Hz$^{-1}$ M$^{-1}_{\odot}$ yr  and low X-ray luminosities of $L_{X,0.2-95{\rm keV}}/SFR < 7.6 \times 10^{39}$ erg  s$^{-1}$  M$^{-1}_{\odot}$ yr. These luminosities correspond to $f_X > 0.25$  and $f_{\rm Radio} <397$  constrained at the 68\% confidence level individually.

The same measurement can be used to constrain the phenomenological synchrotron background model. Here we find that scenarios with a high excess radio background of $A_r > 31$ (where $A_r$ is the amplitude of the excess radio background relative to the CMB  calculated at the reference frequency of 78 MHz assuming $\beta = -2.6$ spectral index), corresponding to 1.6\% of the CMB at 1.42~GHz, are excluded at 68\% confidence. In a similar manner, models with a low X-ray efficiency of $f_X < 0.20$ (corresponding to the X-ray luminosity per SFR of $L_{X,0.2-95{\rm keV}}/SFR < 5.9 \times 10^{39}$ erg s$^{-1}$  M$^{-1}_{\odot}$ yr) are excluded.  
 
Global signal experiments are also delivering constraining data. In disagreement with the EDGES collaboration which reported the puzzling detection,  SARAS2  and SARAS3 experiments put limits on the 21-cm signal from cosmic dawn and the EoR \citep{Singh:2018,Singh22}. Specifically, SARAS2 takes measurements at $z = 7-12$ while SARAS3 probes the 21-cm signal from $z\sim 15-32$. In agreement with HERA, these measurements disfavor a strong radio background at the EoR and weak X-ray heating (see \citet{Bevins_2022} and  \citep{Bevinsprep}). The SARAS experiments probe  a much wider redshift range than HERA thus providing a unique view into the astrophysical processes deep at cosmic dawn.  
 
\subsection{A Stimulating Explanation of the Extragalactic Radio Excess --- Andrea Caputo}\label{Caputo}

Beyond the standard model (BSM) explanations proposed thus far to solve the radio excess background have primarily focused on synchrotron emission from DM annihilation and decay~\citet{CV14,Fornengo11,FL14}. However, these models typically run into several issues, for example they may result in nonsmooth emission or underproduce $T_{\rm exc}$ (here defined as the majority of $T_{\rm BGND}$ in equation (\ref{T_B}) which is in excess of that which would result from known source classes), unless the magnetic fields responsible for synchrotron production have unusual properties~\cite{CV14,FL14}. In~\citet{Caputo:2022npg}, we have proposed a new simple class of experimentally viable new-physics models which can explain the amplitude, power-law dependence and smoothness of $T_{\rm exc}$. These models rely on three basic ingredients: 1) a particle decaying into dark photons $A'$; 2) the presence of a thermal bath of $A'$ which stimulates this decay; and 3) $A'$ resonantly oscillating into radio photons.  A particle physics model that has these three features can generate an RSB with $T_{\rm exc} \propto \omega^{-5/2}$.

There are several parameters that depend on the specifics of the chosen model. In~\citet{Caputo:2022npg} the fiducial model involves a DM axion-like particle (ALP) of mass $m_a$ decaying into an  $A'$ with energy $\omega_{A'} = m_a/2$, in the presence of a thermal bath of  $A'$ which stimulates the decay due to Bose enhancement, leading to a redshift-dependent effective decay lifetime $\tau(z)$
\begin{equation}
	\tau(z) = \tau_{\rm vac} \left[1 + 2 f^{\rm BB}_{A'}(z) \right]^{-1} \,, 
	\label{eq:tau_stimulated}
\end{equation}
with $f^{\rm BB}_{A'} = (e^{\omega_{A'} / T'} - 1)^{-1}$ being the blackbody occupation number of $A'$ with energy $\omega_{A'}$ and temperature $T'$. 

Once the $A'$ are produced, they can then oscillate into ordinary photons via the so called kinetic mixing, $\epsilon$, which enables resonant conversion between the photons and dark photons whenever their masses match $m_{A'}^2 = m_\gamma^2$, where $m_\gamma^2$ is the effective photon plasma mass squared. This quantity is proportional to the free electron number density, $n_{\rm e}$. The converted photons ultimately form the present-day $T_{\rm exc}$. We calculate the sky-averaged conversion probability per redshift $ \langle d P_{A' \to \gamma} \rangle / d z$ taking into account inhomogeneities using the formalism developed in~\citet{Caputo:2020bdy}.

All in all, the number density of produced photons within this mechanisms simply reads 

\begin{equation}
	\frac{d n_\gamma}{d x}(x,z) = \frac{\rho_a(z)}{m_a} \frac{1}{x} \underbrace{\frac{1}{\tau(z_\star)}}_{\propto \, x^{-1}}\underbrace{\frac{1}{H(z_\star)}}_{\propto \, x^{3/2}} \underbrace{\int_z^{z_\star} d z' \frac{\langle d P_{A' \to \gamma} \rangle}{d z'}}_{\propto \, x^{-1}} \,.
	\label{eq:dn_gamma_d_omega_final}
\end{equation}
where $\rho_a(z)$ is the DM density at redshift z, $x \equiv \omega/[T_0(1+z)]$, $z_\star$ is the redshift at which a daughter $A'$ with frequency $\omega$ at redshift $z$ was produced -- $1 + z_\star \equiv \omega_{A'}(1+z)/ \omega$ -- and we assumed all the decays to happen during matter domination (and therefore $H(z_\star) \propto x^{3/2}$). This gives $d n_\gamma / d x \propto x^{-3/2}$, or $T_{\rm exc} \propto \omega^{-5/2}$, the desired frequency dependence.

\begin{figure}[b!]
\includegraphics[width=0.45\textwidth]{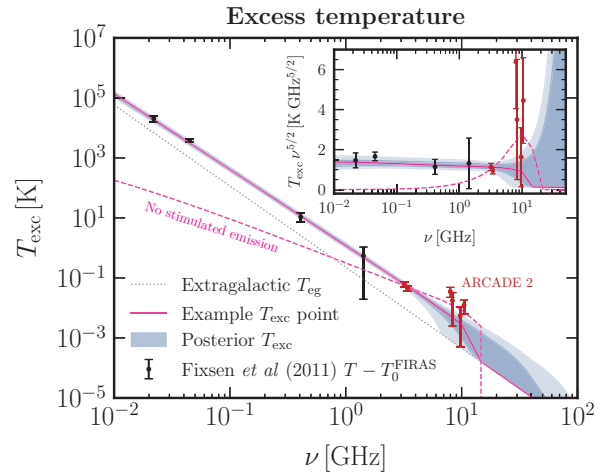}
\caption{Point-wise posterior for $T_{\rm exc}$ within our proposed model, showing the middle-68\% and 95\% regions (dark and light blue regions, respectively). We include the contribution from unresolved extragalactic sources $T_{\rm eg}$~\citep[dashed grey --][]{Gervasi:2008rr} for completeness. The spectrum for a single point in parameter space is shown in pink. Radio data, plotted as $T - T_0^{\rm FIRAS}$, include measurements from ARCADE~2, shown in red~\citep{Fixsen11}, with results from other telescopes shown in black. The pink dashed line shows a fit to the ARCADE~2 data only assuming no stimulated emission.}
\label{fig:T_exc_fiducial}
\end{figure}

Using this simple result we have performed a scan of the model parameter space, individuating the preferred regions which provide a good fit to radio data at all frequencies, while avoiding all other bounds for the considered model. We explored the posterior on the model parameters using nested sampling. Our priors are constructed such that they have zero probability density in parameter regions incompatible with 1) the FIRAS spectral distortion limits of~\citet{Caputo:2020bdy} and 2) DM lifetime limit~\citep{Simon:2022ftd}. Priors on $T_0'$ and $T_0$ are also chosen to account for the number of relativistic degrees of freedom $N_{\rm eff}$ and FIRAS constraints on these parameters, respectively. The posterior on the excess temperature is shown in Fig.~\ref{fig:T_exc_fiducial}, where we also included an irreducible contribution from unresolved extragalactic radio sources~\citep[dashed gray line][]{Gervasi:2008rr}. We find an excellent fit to the data over a wide range of allowed model parameters. 

Finally, as already mentioned, another challenge for models explaining the radio background is its smoothness. There are two possible contributions to the anisotropy produced by our model: 1) decay anisotropy, due to DM density correlations from the point at which $a$ decays, and 2) conversion anisotropy, due to correlations in electron density fluctuations,  $\delta_{\rm e}$, since $m_\gamma^2 \propto n_{\rm e}$. The decay anisotropy was found to exceed the radio anisotropy power spectrum unless the decay happens at $z_\star \gtrsim 5$~\cite{Holder14}, a criterion easily satisfied in the range of parameters providing a good fit. We therefore computed the conversion anisotropy power spectrum by first writing down the two-point correlation function of the conversion probability of $A' \to \gamma$ in two different directions in the sky. Our computation~\citep{Caputo:2022npg} shows that producing a sufficiently smooth radio background is highly plausible within the model, and calls for further studies of the two-point function of $\delta_{\rm e}$ (which for simplicity we modeled as either a Gaussian or a log-normal random field).

Future experiments can probe our model. PIXIE~\citep{PIX} can be sensitive to CMB spectral distortions due to $A'$, and is expected to almost fully cover the 95\% region of our posterior distribution in the $\epsilon$-$m_{A'}$ plane. The thermal population of $A'$ may also lead to a value of $N_{\rm eff}$ that is detectable in future CMB experiments such as CMB-S4~\citep{Abazajian:2019eic}.

\subsection{WIMP Dark Matter and Radio Correlations --- Elena Pinetti}\label{Pinetti}

\begin{figure}[b!]
\includegraphics[width=0.45\textwidth]{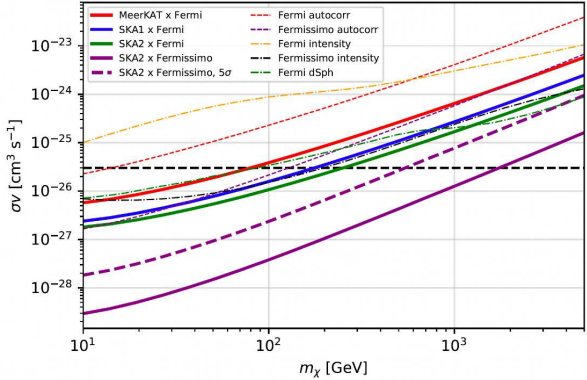}
\caption{Limits on the cross section as a function of the DM mass. Comparison with limits from: autocorrelation of gamma rays (dashed red for Fermi \citet{Fermi_2018}, light purple dashed for Fermissimo), intensity of the gamma-ray flux (yellow for Fermi \citet{Fermi_2015}, black for Fermissimo) and dwarf spheroidal galaxies \citep{Dwarfs_2017}.}
\label{fig:PinettiFig}
\end{figure}

The nature of DM remains a puzzle to date. One enticing possibility is that it is made up of new particles, which may annihilate and produce a huge variety of astrophysical messengers that we can study in order to infer meaningful information on the DM's properties. In \citet{Pinetti_JCAP} we focused on the prompt gamma-ray flux produced by particles with a mass between 10~GeV and 10~TeV, annihilating into $b$ $\bar{b}$ quarks. Astrophysical observations and N-body simulations reveal that DM in the universe is distributed in a hierarchical and anisotropic way. Therefore, we expect that  the electromagnetic signals indirectly produced by DM particles will exhibit a certain degree of anisotropy as well (see \citet{Fornasa_2015} for a review). This is particularly interesting considering that the unresolved gamma-ray background (defined as the total gamma-ray flux minus the Galactic plane and the resolved astrophysical sources) is observed to be anisotropic \citep{Fermi_2012} . In \citet{Pinetti_JCAP} we investigated whether there is a correlation between the anisotropies of the gamma-ray sky and a DM signal, taking into account also the presence of the astrophysical background due to the unresolved astrophysical sources (mainly star-forming galaxies and active galactic nuclei). We employed the cross-correlation technique, which is a powerful method to estimate the correlation between the fluctuations of a gravitational tracer - which is a manifestation of the existence of DM - and an electromagnetic signal - which is a byproduct of the exotic nature of DM.
If we find a positive signal in this cross-correlation channel, we can argue that DM is made up of exotic physics and is not, for instance, the result of an alternative theory of gravity. For the first time we forecasted the cross-correlation signal between the unresolved gamma-ray background and the distribution of neutral hydrogen, measurable with radio telescopes. Neutral hydrogen atoms produce the 21cm line via the spin-flip transition from a higher energy level to the ground state. This emission line acts as a promising gravitational tracer, notably in view of the next-generation experiment SKA, which will be the world's largest radio telescope, with over a square kilometer of collecting area \citep{SKA_2020}. The SKA is currently under construction, while its precursors are already taking data, including MeerKAT which is relevant for this analysis. These radio detectors will be used for 21cm line intensity mapping: a cutting-edge technique, which consists of measuring the integrated emission of a the 21cm line, originating from the intergalactic medium and galaxies \citep{Bull_2015}. The main advantage of 21cm intensity mapping is the great redshift resolution, which is a key factor for DM searches, where we expect the low-redshift structures to dominate our DM signal, while the contribution of the unresolved astrophysical sources is peaked at higher redshift (typically between 0.5 and 1). Regarding the gamma-ray emission, we considered Fermi-LAT as the benchmark detectors. We also take into account a next-generation gamma-ray telescope, which we called Fermissimo, with improved exposure and angular resolution. The angular power spectrum is the statistical tool employed to estimate the cross-correlation signal:
\begin{equation}
    C_\ell^{\gamma \times HI} = \int d z \, \frac{c}{H(z) \chi^2(z)} \, W_\gamma(z) \, W_{HI}(z) \, P_{\gamma \times HI}\left(k=\frac{\ell}{\chi(z)} \right)
\end{equation}
where  $W_\gamma(z)$ and $W_{HI}(z)$ are the two window functions associated to the gamma-ray flux and 21cm line emission, respectively, and they encoded the information on the redhisft evolution of the observables; while $P_{\gamma \times HI}$ is the Fourier power spectrum. The variance on the $C_\ell^{\gamma \times HI} $ has been derived under the hypothesis of Gaussianity. The angular power spectrum and its variance have been employed to derive the forecasted bounds on the DM parameter space and determine the opportunities for DM searches offered by radio correlations with next-generation radio telescopes. Fig.~\ref{fig:PinettiFig} illustrates the $2 \sigma$ limits on the velocity-averaged annihilation cross section (vertical axis) as a function of the DM mass (horizontal axis) for different combination of detectors, as indicated in the legend (solid lines). Other constraints are also shown for comparison (see caption for details). The figure shows that the configurations MeerKAT $\times$ Fermi-LAT and SKA1 $\times$ Fermi-LAT are already competitive with current constraints, while Fermissimo would allow to investigate DM masses up to few TeV, with a $5 \sigma$ detection reach possible for thermal particles with masses up to 400 GeV.

\subsection{Dark Matter Constraints from Planck Observations of the Galactic Polarized Synchrotron Emission  --- Alessandro Cuoco}

High-energetic cosmic-ray (CR) electrons and positrons ($e^\pm$) can be of astrophysical origin or be produced by the annihilation or decay of DM particles in the halo of DM particles surrounding the Galaxy.  They can then propagate in the interstellar Galactic magnetic field (GMF), and produce radio and microwave synchrotron emission.   The potential DM synchrotron signal in the Galactic halo and in extragalactic targets has been studied using radio and microwave surveys, such as WMAP and \textit{Planck}, \citep{Delahaye:2011jn,Egorov:2015eta,Cirelli:2016mrc,Fornengo11,Hooper14,Fornengo14}.  These DM studies have typically focused on the total intensity, i.e., the Stokes parameter $I$. However, synchrotron emission from $e^\pm$ is partially linearly polarized, and a  polarization signal (i.e., Stokes $P$) is thus expected.  In \citet{Manconi:2022vci}, we have exploited the \textit{Planck} polarization maps in order to constrain the Galactic DM signal.  In the following we will briefly summarize our findings, referring to \citet{Manconi:2022vci} for a  more detailed description.
 
To constrain the DM signal we will use intensity and polarization maps from the latest \textit{Planck} data release \citep{Planck:2018nkj}, at frequencies of 30, 44 and 70~GHz.  Furthermore, we also build the related error maps using a procedure which estimates the variance of the signal in each pixel using the neighboring pixels.
At these frequencies the emission is dominated by the Galactic foregrounds rather than by the CMB.  Nonetheless, we do not attempt to model and subtract them.  Instead, we  will derive conservative constraints requiring that the DM signal does not exceed the observed intensity or level of polarization.

We consider WIMPs in the mass range between $5$~GeV and $1$~TeV for annihilation into three representative channels:  $\tau^+ \tau^-$ and $\mu^+ \mu^-$, producing a hard $e^\pm$ spectrum, and $b\bar{b}$, giving a softer spectrum.  To propagate the $e^\pm$ in the Galaxy and derive the all-sky synchrotron maps from DM annihilations we use the \texttt{GALPROP} code version v54r2766 \footnote{Publicy available at \url{https://gitlab.mpcdf.mpg.de/aws/galprop}} as adapted  in Ref.~\citep{Egorov:2015eta} \footnote{Publicy available at \url{https://github.com/a-e-egorov/GALPROP_DM}}. 

The main systematic uncertainty of the analysis is expected to come from the modeling of the GMF, because it is not well constrained~\citep{Jaffe:2019iuk}.  The GMF is known to have at least two components:  an isotropic turbulent, random field and a large-scale regular one.  To gauge the uncertainties related to the GMF, we consider the following three models:  The Sun+10 model~\citep{Sun08,Sun:2010sm}, the Psh+11 model~\citep{Pshirkov:2011um}, and the JF12 model proposed by Jansson \& Farrar \citep{Jansson:2012pc,Jansson:2012rt}.  These three models differ in both the regular and the random MF components.  It's important to note that the intensity and polarization signal depend differently on the MF properties.  Specifically, while the intensity depends on the total (random+ordered) MF strength, polarization only depends on the regular component of the field.  

\begin{figure}[t]
	\centering
	\includegraphics[width = 0.49\textwidth]{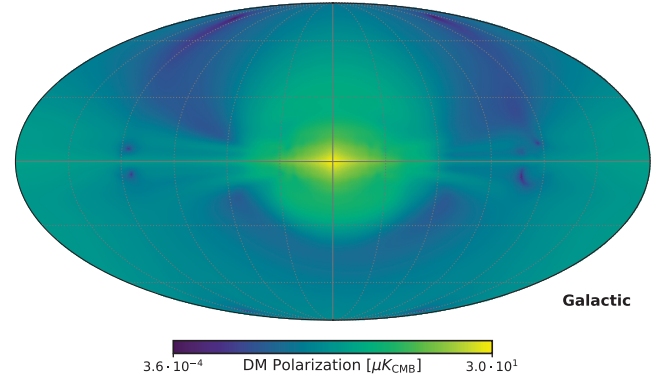}
	\caption{Example morphology of the DM  synchrotron polarization amplitude  at $30$~GHz for $m_{\rm DM}=50$~GeV annihilating in $\mu^+ \mu^-$ pairs with a thermal cross section,  and  Psh+11  GMF model. }
	\label{CuocoAFig}
\end{figure}

We show in Fig.~\ref{CuocoAFig} an example of the morphology of the synchrotron polarization amplitude  from DM at $30$~GHz for $m_{\rm DM}=50$~GeV annihilating in $\mu^+ \mu^-$ pairs with a thermal cross section,  and  for the Psh+11  GMF model.  For each DM mass and annihilation channel, we compute the DM intensity and polarization map  and derive upper bounds on the  annihilation cross section by requiring that the DM emission does not exceed the observed \textit{Planck} signal \textit{plus} the error estimated before.  The final quoted upper limits are given by the most constraining pixel. 

\begin{figure*}[t]
		\includegraphics[width = 0.49\textwidth]{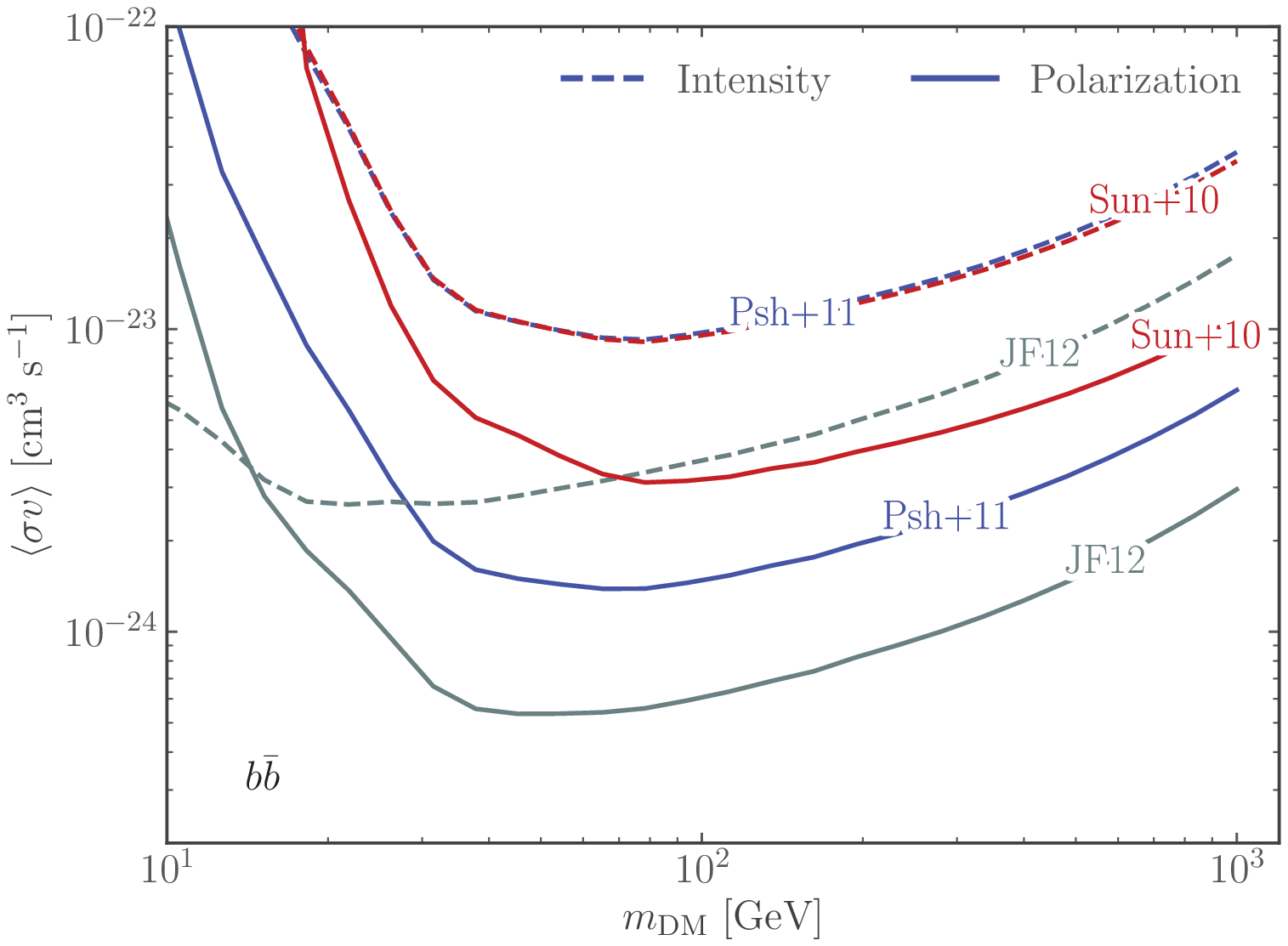}
	\includegraphics[width = 0.49\textwidth]{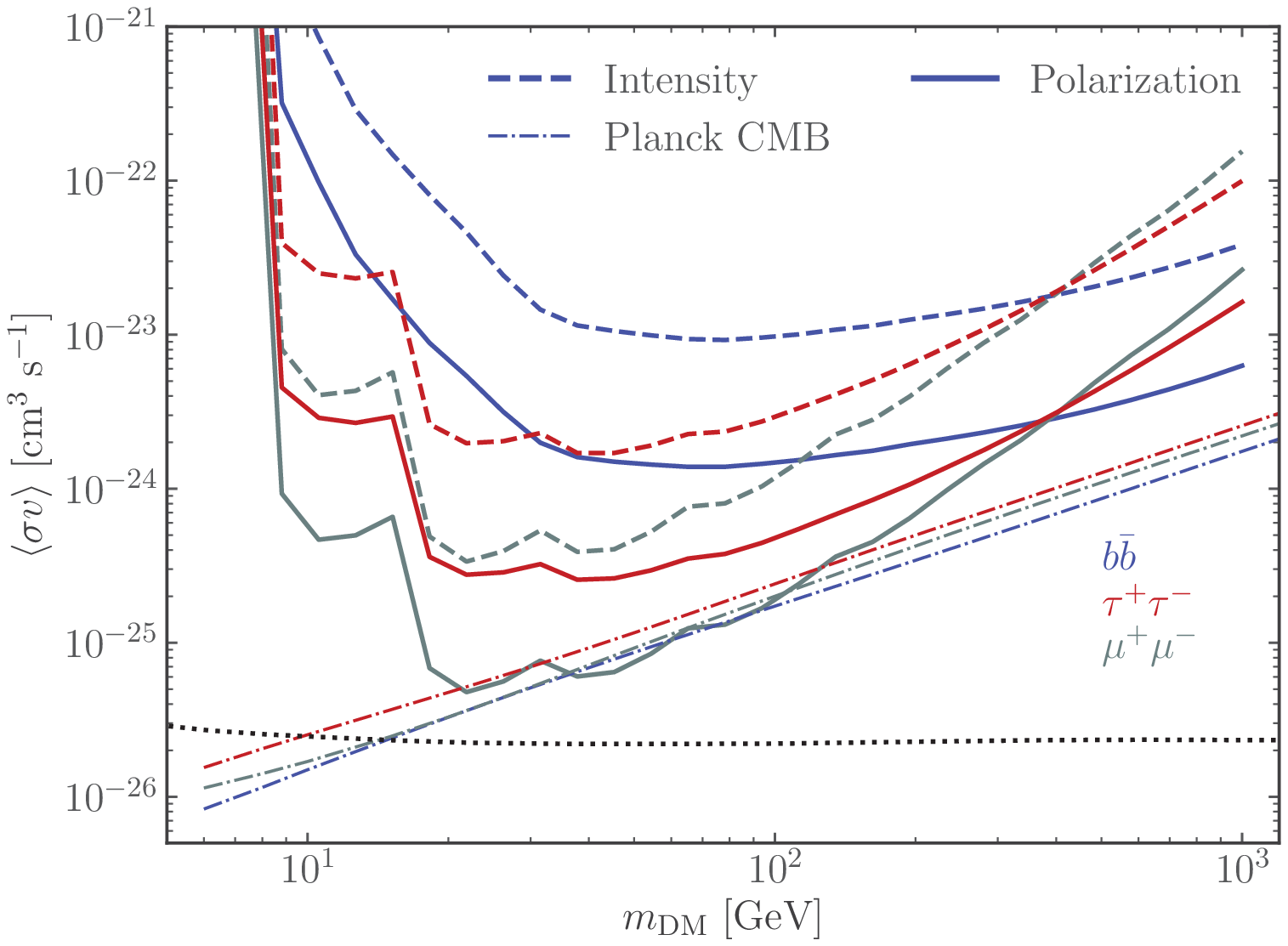}
	\caption{ Upper limits on the DM annihilation cross section from the  intensity and polarization  data at $30$~GHz.
	Left: effect of the GMF modeling for the $\bar{b} b$ channel. Right: constraints for different annihilation channels for the Psh+11 GMF. Also shown for comparison are the constraints obtained from \textit{Planck} CMB data \citep{Planck:2018vyg}. 
	 }
	\label{CuocoBCFig}
\end{figure*}

The DM constraints obtained for the $\bar{b} b$ channel for intensity and polarization are illustrated in Fig.~\ref{CuocoBCFig} (left) for different GMF models. This illustrates the main result of the analysis, namely the fact that polarization limits are about a factor of ten more constraining than the ones derived from intensity maps. As expected, however, the GMF provides a significant uncertainty also of about one order of magnitude.  Nonetheless, polarization is always more constraining than intensity, independently of the GMF model.  Further systematic uncertainties related to the choice of the propagation setup or the DM radial profile are discussed  in \cite{Manconi:2022vci}. 

The upper limits for the three annihilation channels considered are shown in Fig.~\ref{CuocoBCFig} (right), for a fixed choice of Psh+11 GMF.  For the $\mu^+ \mu^-$ channel the polarization upper limits at several tens GeVs are close to the thermal relic value  \citep{Steigman:2012nb} and
competitive with \textit{Planck} CMB constraints~\cite{Planck:2018vyg}.

The stronger DM constraints from polarization come mainly from two effects.  First, the observed intensity and polarization emissions have significantly different morphologies, with polarization presenting filaments, or arms, and leaving interarm regions with low background very close to the Galactic center, where the DM signal is highest.  On the contrary, the observed intensity is quite uniform and the S/N for DM peaks further away from the Galactic center.  Second, polarization presents a lower overall background level.

In summary, we have derived, for the first time, limits on the DM annihilation cross section from
\textit{Planck} observations of the polarized synchrotron  emission, finding constraints about one order of magnitude stronger than the ones derived from intensity.  The constraints can be further improved by a proper modeling and removal  of the astrophysical synchrotron signal.  We leave this assessment to future work.

\section{Discussion}\label{conc}

A main consensus of the workshop was that there are encouraging prospects for near-term analysis projects related to both i) the radio~SZ effect (discussed in \S \ref{Holder}) and ii) cross-correlation analyses.  These may either not require much in the way of new observations or may require new observations that are feasible in scope.

As discussed in \S \ref{Holder} a positive detection of the radio~SZ effect would confirm the RSB as extragalactic.  Additionally redshift tomography would be possible utilizing clusters of different redshifts, allowing a potential constraint to the redshift of origin of the RSB.  There may already be sufficient radio observations of clusters in the literature from LOFAR and/or MeerKAT.  In the case of LOFAR,  where the radio SZ effect would manifest as an increase in emission which would have to be separated from emission due to sources in a cluster itself, these could be targeted observations of clusters \citep[e.g.][]{Savini19} or clusters surreptitiously seen in surveys such as the LOFAR Two-metre Sky Survey \citep[LoTSS --][]{Shimwell17}.  In the case of MeerKAT the MeerKAT Cluster Legacy Survey \citep[MGCLS--][]{Knowles22} has detected over 100 clusters at 8'' resolution in L band (900~--~1670~MHz) where the radio~SZ effect would manifest as a decrement.  

For cross-correlation analyses, in addition to the observations of \citet{LF21} discussed in \ref{Heston}, there are also legacy LOFAR observations of other fields and the possibility of targeting additional fields.  These radio observations could be cross correlated with available maps at other wavelengths which trace distinct structures in the universe.  These include CMB lensing \citep[e.g. Planck--][]{Planck20} which traces the overall matter distribution, catalogs in optical which trace the galaxy distribution, and X-rays which trace the black hole distribution.  The latter is important for potentially constraining PBH origin scenarios such as in \S \ref{Cappeluti} and \S \ref{Mittal}.

Another consensus of the workshop was that the 310~MHz absolute map project discussed in \S \ref{GBTsec} and \S \ref{Bordenave} should be prioritized over the other potential measurements discussed at the first RSB workshop.  These other proposed measurements were i) a targeted one at 120~MHz and ii) a measurement in the GHz range with greater sky coverage than ARCADE~2.  However we did agree that a future measurement to determine whether the synchrotron spectrum hardens at frequencies $>$~10~GHz would be valuable, in particular for discriminating between various proposed origin scenarios for the RSB.  Such a measurement would likely require a future space mission, although the Tenerife Microwave Spectrometer \citep[TMS --][]{TMS} may prove useful in this regard.  

We also discussed the need to understand more about the small-scale structure of the Galactic magnetic field, which is necessary for investigations that use the observed polarization information to constrain a potential Galactic origin for the RSB such as in \S \ref{Kogut}.  There are already comprehensive maps of Faraday rotation in the literature \citep[e.g.][]{Oppermann12} which cannot be explained with current models of the Galactic magnetic field, so larger-scale models of the Galactic magnetic field are necessary.  We look forward to the publication of polarization maps from C-BASS (discussed in \S \ref{Leahy} and \S \ref{Harper}) and the Australia Square Kilometer Array Pathfinder (ASKAP) polarization Sky Survey of the Universe's Magnetism \citep[POSSUM -- e.g.][]{Anderson21}.

Another point raised, which was also emphasized at the first RSB workshop, is that proposed origin scenarios should be testable with observations.  These observations could potentially be in the radio and/or some other waveband(s), at any angular scale(s), in intensity and/or polarization, and either new or available in the literature.

The workshop and its format were unanimously declared a success by the participants, and, along with the consensus on the issues discussed above, there was agreement on the importance of the topic and of another dedicated meeting in the time span of 4--5 years.

\acknowledgments

We thank the INFN and University of Torino for the financial support to the meeting, and the Hotel Barolo and Brezza restaurant and winery for their hospitality.  J.S. acknowledges support from a University of Richmond College of Arts and Sciences Sabbatical Fellowship. M.R. and E.T. acknowledge funding from the PRIN research grant No. 20179P3PKJ and the ``Department of Excellence" grant funded by MIUR, and from the research grant TAsP funded by INFN.  S.H. acknowledges support from NSF grant No.~PHY-1914409. E.P. acknowledges the support from the Ministero degli Affari Esteri della Cooperazione Internazionale - Direzione Generale per la Promozione del Sistema Paese Progetto di Grande Rilevanza ZA18GR02.

\end{document}